\documentclass[lettersize,journal]{IEEEtran}
\usepackage{amsmath,amsfonts}
\usepackage{algorithmic}
\usepackage{algorithm}
\usepackage{array}
\usepackage{textcomp}
\usepackage{stfloats}
\usepackage{url}
\usepackage{verbatim}
\usepackage{graphicx}
\usepackage{multirow}
\usepackage{booktabs}
\usepackage{makecell}
\usepackage{xcolor}
\usepackage{hyperref}
\usepackage{subcaption}
\usepackage[numbers, sort&compress]{natbib}
\usepackage{stmaryrd}

\makeatletter
\newcommand\footnoteref[1]{\protected@xdef\@thefnmark{\ref{#1}}\@footnotemark}
\makeatother

\def\BibTeX{{\rm B\kern-.05em{\sc i\kern-.025em b}\kern-.08em
    T\kern-.1667em\lower.7ex\hbox{E}\kern-.125emX}}
\usepackage{balance}

\DeclareMathOperator{\relu}{ReLU}
\DeclareMathOperator{\sinc}{sinc}

\newcommand{\revision}[1]{\textcolor{black}{{#1}}}
\newcommand{\rerevision}[1]{\textcolor{black}{{#1}}}
\newcommand{\flatMapsto}{\mathrel{\vcenter{\hbox{\rule{0.4pt}{0.48ex}}}\mkern-6.5mu\Rightarrow}}

\begin{document}

\title{Aliasing-Free Neural Audio Synthesis}

\author{Yicheng Gu, \textit{Student Member, IEEE}, Junan Zhang, Chaoren Wang, Jerry Li, \\ Zhizheng Wu, \textit{Senior Member, IEEE}, Lauri Juvela, \textit{Member, IEEE}

\thanks{\vspace*{-14pt}}
\thanks{Received 23 December 2025; revised 2 April 2026 and xx May 2026. This work was supported in part by the Aalto University School of Science ``Science-IT'' project, and in part by the EuroHPC Joint Undertaking, which awarded this project access to the EuroHPC supercomputer LUMI, hosted by CSC and the LUMI consortium through a EuroHPC Regular Access call.}
\thanks{Jerry Li is with the Spellbrush, Akihabara, Tokyo 101-0021, Japan (e-mail: jerry@sizigistudios.com).}
\thanks{Lauri Juvela is with the Acoustic Laboratory, Department of Information Communications Engineering, Aalto University, Espoo 02150, Finland (e-mail: lauri.juvela@aalto.fi).}
\thanks{Junan Zhang, Chaoren Wang, and Zhizheng Wu are with the School of Data Science, The Chinese University of Hong Kong, Shenzhen, Guangdong 518172, China (e-mail: junanzhang@link.cuhk.edu.cn; chaorenwang@link.cuhk.edu.cn; wuzhizheng@cuhk.edu.cn).}
\thanks{Yicheng Gu is with the Spellbrush, Akihabara, Tokyo 101-0021, Japan, also with the Acoustic Laboratory, Department of Information Communications Engineering, Aalto University, Espoo 02150, Finland, and also with the School of Data Science, The Chinese University of Hong Kong, Shenzhen, Guangdong 518172, China (e-mail:yichenggu@link.cuhk.edu.cn).}
}




\maketitle
 
\begin{abstract}


\revision{In neural audio synthesis, neural vocoders and codecs are models that reconstruct waveforms from acoustic and latent representations, which are essential to the resulting audio quality.}
\revision{While current models are capable of generating perceptually natural speech, they still struggle with high-fidelity music and singing voice synthesis, as severe aliasing artifacts are introduced by non-linear activation functions and upsampling layers in existing architectures.}
\revision{Although various anti-aliasing techniques have been proposed in digital signal processing, their integration into neural vocoders and codecs remains under-explored.}
\revision{This paper incorporates differentiable anti-aliasing techniques into the activation and upsampling modules to bridge this gap, and thus presents Pupu-Vocoder and Pupu-Codec.}
\revision{We build a test signal benchmark to evaluate the anti-aliased modules, and validate our proposed models on speech, singing voice, music, and audio.}
\revision{Experimental results show that Pupu-Vocoder and Pupu-Codec outperform existing systems on singing voice, music, and audio, while achieving comparable performance on speech.}
\rerevision{Demos, codes, and checkpoints are available at \url{VocodexElysium.github.io/AliasingFreeNeuralAudioSynthesis/}.}

\end{abstract}

\begin{IEEEkeywords}
neural codec, neural vocoder, speech synthesis, singing voice synthesis, audio synthesis, DDSP, anti-aliasing.
\end{IEEEkeywords}
\vspace{-11pt}

\section{Introduction}
\vspace{-3pt}
\IEEEPARstart{E}{xisting} audio generation systems typically consist of two stages. Firstly, an acoustic model~\cite{maskgct, vevo, valle} converts \revision{task-specific inputs (e.g., text in text-to-speech synthesis)} into an intermediate representation. Then, a decoder model, often referred to as a vocoder, reconstructs the waveform from it. Among different types of vocoders, the neural network-based ones~\cite{WaveRNN, lpcnet, WaveNet, wavefm, WaveFlow, WaveGlow, WaveGrad, DiffWave, fregrad, ddsp-vocoder, glotnet, sawsing, bigvgan, HiFiGAN, evagan} are essential due to their superior synthesis quality compared with the DSP-based ones~\cite{phase-vocoder, straight, world}. Historically, the development of neural vocoders has primarily focused on time-domain models that generate waveforms by upsampling acoustic representations without explicitly modeling the phase. These models include flow-based~\cite{wavefm, WaveFlow, WaveGlow}, diffusion-based~\cite{WaveGrad, DiffWave, fregrad}, auto-regressive-based~\cite{WaveRNN, lpcnet, WaveNet}, differentiable digital signal processing (DDSP)-based~\cite{ddsp-vocoder, glotnet, sawsing}, and generative adversarial network (GAN)-based vocoders~\cite{bigvgan, evagan, HiFiGAN}.

Among these various approaches, GAN-based vocoders are extensively studied due to their faster inference speed and higher synthesis quality. Following these developments, GAN-based neural audio codecs~\cite{encodec, dac, speechtokenizer} have also been proposed in recent years. The main idea of the neural audio codec is to decompose the continuous intermediate representation into discrete tokens via an encoder with \revision{vector~\cite{vq, vqvae, soundstream} or scalar~\cite{FSQ-1, FSQ-2, fsq} quantization}, which can later be transformed back and fed to a vocoder-like decoder to obtain the sound. Such a discretization \revision{facilitates the use of language models, resulting in state-of-the-art (SOTA) performance~\cite{maskgct, vevo, valle}}.

Although these upsampling-based time-domain models can generate perceptually natural sound, recent studies suggest that their synthesis fidelity remains limited due to aliasing artifacts brought by inadequately designed model architectures~\cite{ismir-upsample, interpolation, upsample-filter, wavehax}. Specifically, the unconstrained nonlinear activation generates infinitely many harmonics that exceed the Nyquist frequency, resulting in ``folded-back'' aliasing artifacts~\cite{bigvgan, wavehax}. The widely used upsampling layer, ConvTranspose, copies the mirrored low-frequency parts to fill the empty high-frequency region, resulting in ``mirrored'' aliasing artifacts~\cite{ismir-upsample, interpolation, upsample-filter}. Meanwhile, the combination of its inherent periodicity and the mirrored DC bias also introduces ``tonal artifact''~\cite{ismir-upsample, interpolation, upsample-filter}, resulting in constant-frequency ringing. Several recent works have been proposed to address these issues, but at the cost of either synthesis speed~\cite{bigvgan} or quality degradation~\cite{ismir-upsample, interpolation, upsample-filter}. Others attempt to use the aliasing-free time-frequency (TF) domain model as an alternative~\cite{stftcodec, vocos, wavehax}, which generates TF representations (TFRs) and uses their inverse transforms to obtain the audio; \revision{but their synthesis quality remains limited due to the difficulty in explicitly modeling the phase~\cite{stftcodec}.}

This paper aims to address these problems inherent in activation and upsampling modules from a signal processing perspective. Specifically, we apply oversampling and anti-derivative anti-aliasing (ADAA) techniques~\cite{adaa-nonlinear, adaa-origin, adaa-wavedigital, adaa-state, adaa-iir, adaa-rnn} to the activations to obtain their anti-aliased form, and replace the ConvTranspose layer with resampling to avoid ``tonal artifact'' and eliminate aliased components. Based on our proposed anti-aliased modules, we introduce Pupu-Vocoder and Pupu-Codec, and release high-quality pre-trained checkpoints to facilitate audio generation research. We build a test signal benchmark to illustrate the effectiveness of the anti-aliased modules, and conduct experiments on speech, singing voice, music, and audio to validate our proposed models. Experimental results confirm that our lightweight Pupu-Vocoder and Pupu-Codec models can easily outperform existing systems on singing voice, music, and audio, while achieving comparable results on speech. Beyond audio synthesis, we emphasize that our method is domain-agnostic, making it applicable to other domains where aliasing is a concern, such as image generation.

\section{Theoretical Background}
\label{sec:analysis}

\begin{figure*}[t]
    \centering
    \includegraphics[width=0.95\textwidth]{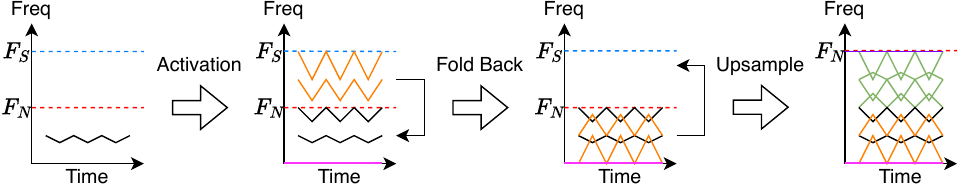}
    \caption{Illustration of different aliasing artifacts brought by the activation functions and upsampling layers. $F_S$ denotes the sampling rate, $F_N$ denotes the Nyquist frequency. The colored contours represent ``folded-back'' aliasing artifacts (orange), ``mirrored'' aliasing artifacts (green), DC bias (pink), and ``tonal artifact'' (purple), respectively. \revision{For simplicity, only the one-sided positive frequency spectrum of the signal is illustrated, omitting the negative frequencies dictated by Hermitian Symmetry.}}
    \vspace{-8pt}
    \label{fig:aliasing}
\end{figure*}

In this section, we discuss the related work and outline the theoretical background of the ``folded-back'' and ``mirrored'' aliasing artifacts, as well as the ``tonal artifact'' introduced by the activation functions and upsampling layers.

\subsection{Artifacts due to Non-Linear Activation Functions}

Applying an unconstrained activation to a discrete signal will generate an infinite number of harmonics that exceed the Nyquist frequency. According to the Nyquist–Shannon sampling theorem~\cite{aliasing-1}, these harmonics will then be ``folded-back'' and become aliasing artifacts, as illustrated in Figure~\ref{fig:aliasing}.

Following Wavehax~\cite{wavehax}, we take the ReLU activation as an example to better illustrate the idea. Suppose the input signal is a sine wave with angular frequency $\omega$ for continuous time $t \in R$, after applying the ReLU activation function, the resulting signal's Fourier expansion becomes:
\begin{equation}
    \relu(\sin(\omega t)) = \frac{1}{\pi} + \frac{\sin(\omega t)}{2} - \sum_{k=1}^{\infty} \frac{2\cos(2k \omega t)}{\pi(2k - 1)(2k + 1)},
\end{equation}
where the last term induces an infinite amount of harmonics. The frequency components higher than the Nyquist frequency, where$\frac{k \omega }{\pi} > \frac{F_{N}}{2}$, would become the aliasing artifacts.

To address this issue, StyleGAN 3~\cite{stylegan3} and BigVGAN~\cite{bigvgan} utilize the oversampling technique to temporarily increase the Nyquist frequency before applying the activation, thus reducing the amount of aliased components. To implement such a technique, upsampling the signal with a low-pass filter before applying the activation, and then downsampling it back with a low-pass filter to remove the extra frequency region, as:
\begin{equation}
\begin{split}
    \hat{x} &= \text{lowpass}(\text{upsample}(x, c), F_{N}), \\
    y &= \text{downsample}(\text{lowpass}(f(\hat{x}), F_{N}), c),
\end{split}
\end{equation}
where $x$ is the input signal, $\hat{x}$ is the upsampled signal, $y$ is the output signal, $f(\cdot)$ is the activation function, $F_{N}$ is the Nyquist frequency which is the half of the sampling rate $F_{S}$, $\text{lowpass}(x, F_{N})$ represents low-pass filtering with a cut-off frequency of $F_{N}$, $\text{upsample}(x, c)$ and $\text{downsample}(x, c)$ denotes upsampling or downsampling the input signal by a factor of $c$. To effectively eliminate the aliased components, an oversampling factor of 4 or 8 is typically required~\cite{adaa-nonlinear}, which makes the signal excessively long and thus incompatible with deep learning applications due to GPU memory constraints.

To avoid these issues, several more advanced anti-aliasing techniques have been proposed. For instance, a harmonic mixed model can be used for polynomial nonlinearities~\cite{harmonic-mixer}, or a non-linearity can be approximated using a filter bank model~\cite{filter-bank}. However, such methods are usually complex and limited to a few function types, excluding the widely used activations. Recently, a simple yet effective method called ADAA~\cite{adaa-nonlinear, adaa-origin, adaa-wavedigital, adaa-state, adaa-iir, adaa-rnn} has been proposed. Its main idea is to convert the discrete signal to a continuous one before applying the activation function. Since the signal is continuous, there are no sampling frequency constraints, and therefore, no aliasing artifacts. \revision{Such a signal can then be low-pass filtered to remove the extra frequencies and discretely sampled back.} \revision{Namely, for continuous time $t \in [0, n]$ and a signal $x$, we have:
\begin{equation}
    y_t = f(\widetilde{x}(t))
\end{equation}
where $\widetilde{x}$ is a continuous-time reconstruction of $x$. We use linear interpolation to obtain $\widetilde{x}$ following~\cite{adaa-origin}, as:
\begin{equation}
\begin{split}
\label{eq:x-continuous}
    \widetilde{x}(t) = 
    \begin{cases}
        x_0 + \tau (x_1 - x_0), & \text{if } 0 \leq t < 1 \\
        \quad \quad \quad \quad \quad \quad \quad\vdots & \\
        x_{n - 1} + \tau (x_n - x_{n - 1}), & \text{if } n - 1 \leq t < n
    \end{cases}
\end{split}
\end{equation}
where $\tau = t - \lfloor t \rfloor$ is a time variable that runs $0...1$ between each sample.} \revision{Applying the activation $f(\cdot)$ to the signal $\widetilde{x}$, followed by low-pass filtering with a filter kernel $h(\cdot)$ and then discretely sampling the resulting signal back, gives:
\begin{equation}
\begin{split}
\label{eq:adaa-before}
    y_t = \int_{-\infty}^{\infty}h(u)f(\widetilde{x}(t - u))du,
\end{split}
\end{equation}
where we define $h(\cdot)$ as a rectangular filter kernel with unit width for future computation following~\cite{adaa-origin}, as:
\begin{equation}
\begin{split}
    h(t) = 
    \begin{cases}
        1, & \text{if } 0 \leq t \leq 1 \\ 
        0, & \text{otherwise} \\
    \end{cases}
\end{split}
\end{equation}
which, following the derivation process in~\cite{adaa-origin}, can be reduced to a closed-form expression as follows:
\begin{equation}
\label{eq:adaa}
    y_t = \frac{F(x_t) - F(x_{t - 1})}{x_t - x_{t - 1}},
\end{equation}
where $F(\cdot)$ is the first order anti-derivative of $f(\cdot)$. To avoid numerical instabilities when $x_t$ is close to $x_{t-1}$, a threshold-based fallback mechanism is further utilized, as:
\begin{equation}
    y_t = f\left(\frac{x_t + x_{t - 1}}{2}\right), \quad \text{if } |x_t - x_{t - 1}| < \epsilon
\end{equation}
where $\epsilon$ is a small tolerance value often set to be $1e-5$.}

\begin{figure*}[t]
    \centering
    \includegraphics[width=0.95\textwidth]{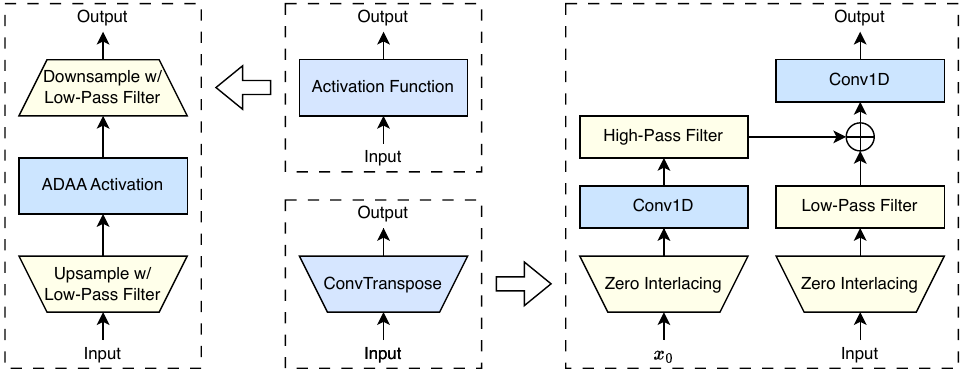}
    \caption{\revision{The main idea of our proposed anti-aliased activation function and upsampling layer. ``$\flatMapsto$'' means ``replacing the original problematic model architecture with our proposed anti-aliased modules''.} $x_0$ is the latent representation obtained from the first Conv1D layer in the decoder. We employ a resampling layer (zero-interlacing + low-pass filtering) with a noise-like, high-pass filtered deterministic prior, obtained from the zero-interlaced $x_0$, to replace the problematic ConvTranspose layer. Additionally, we utilize an oversampled ADAA activation function to replace the original unconstrained one.}
    \label{fig:method}
\end{figure*}

\begin{figure}[t]
    \centering
    \begin{subfigure}[b]{0.49\columnwidth}
         \centering
         \includegraphics[width=\columnwidth]{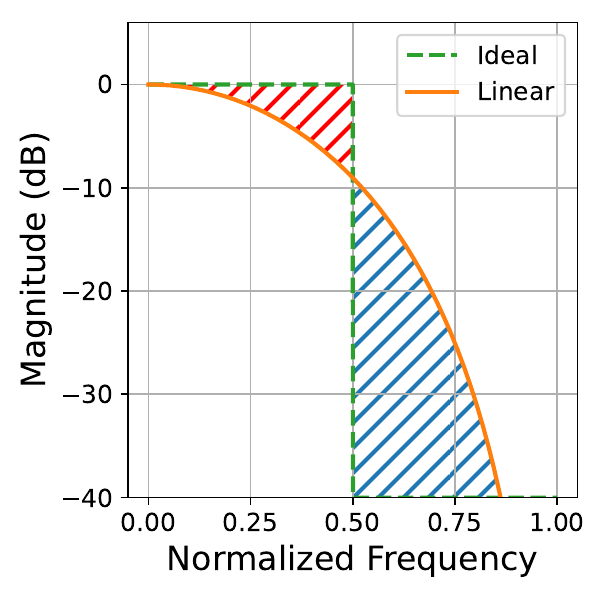}
         \caption{Linear Interpolation}
         \label{fig:linear-response}
    \end{subfigure}
    \hfill
    \begin{subfigure}[b]{0.49\columnwidth}
         \centering
         \includegraphics[width=\columnwidth]{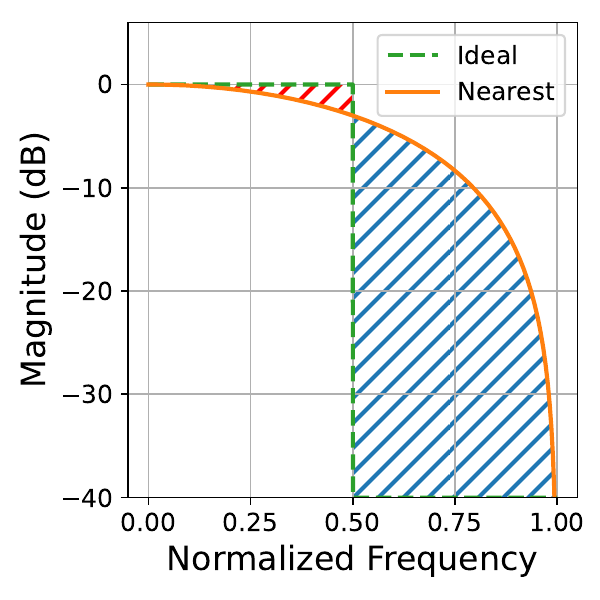}
         \caption{Nearest Interpolation}
         \label{fig:nearest-response}
    \end{subfigure}
    \caption{Illustration of the equivalent filter frequency responses of linear and nearest interpolations. The orange contour and green dashed line represent the actual and ideal frequency responses, respectively. The red hatched area indicates the attenuation of the valid frequency region we want to retain, while the blue hatched area highlights the residual ``mirrored'' aliasing artifacts that the interpolation layer fails to remove.}
    \label{fig:filter-response}
\end{figure}

\revision{Within the broader ADAA framework, the continuous-time reconstruction and the low-pass filter can be any functions tailored to different purposes~\cite{adaa-nonlinear, adaa-origin, adaa-wavedigital, adaa-state, adaa-iir, adaa-rnn}. However, to ensure compatibility with deep learning architectures, we specifically adopt linear interpolation in Equation~\eqref{eq:x-continuous} and a rectangular kernel in Equation~\eqref{eq:adaa-before}, following the original paper~\cite{adaa-origin}. To further improve computational efficiency and maintain a continuous gradient flow during training, we carefully chose SnakeBeta~\cite{snake} as the activation function to eliminate the need for the threshold-based fallback mechanism, as its derived closed-form contains no denominator terms and thus ensures numerical stability, which we will show in Section~\ref{sec:method}.}

\subsection{Artifacts due to Upsampling Layers}

The widely used upsampling layer, ConvTranspose, zero-interlaces the input signal and then applies a convolutional layer. \revision{Fundamentally, this time-domain operation (via zero-interlacing) mathematically corresponds to frequency-domain expansion and spectral replication~\cite{discrete}.} Therefore, in the frequency domain, as illustrated in Figure~\ref{fig:aliasing}, this process can be viewed as copying the mirrored low-frequency parts to fill the empty high-frequency region, resulting in ``mirrored'' aliasing artifacts. The ConvTranspose layer also suffers from the ``tonal artifact'', manifested as constant-frequency ringing, as shown in Figure~\ref{fig:aliasing}. Such a phenomenon originates from two sources. Firstly, the DC bias introduced by non-linear activations or network bias parameters is mirrored into high-frequency bands; meanwhile, the computational process of ConvTranspose has an inherent periodicity due to its fixed stride and shared weights, introducing constant-frequency ringing at the same locations as the mirrored DC bias~\cite{interpolation}. 

To address these problems, a low-pass filter can be adapted after the ConvTranspose layer to eliminate the aliased parts. To compensate for the training instability brought by the filter, a noise-like, high-pass filtered deterministic prior can be utilized to fill the empty high-frequency region~\cite{upsample-filter}. This strategy effectively removes the ``mirrored'' aliasing artifacts, but leaves the ``tonal artifact'' unaddressed. To resolve the ``tonal artifact'', previous studies~\cite{ismir-upsample, interpolation} propose to replace the problematic ConvTranspose layer with the linear and nearest interpolations, since they do not exhibit inherent periodicity and their operations are equivalent to low-pass filtering, which can simultaneously remove the mirrored DC bias in the high-frequency region. However, such a replacement does not effectively eliminate the ``mirrored'' aliasing artifacts, and it will also introduce ``filter artifact'' due to their poor filter frequency responses, resulting in a degradation of quality. Specifically, the linear and nearest interpolations are equivalent to convolving the signal with low-pass filter kernels, as:
\begin{equation}
\begin{split}
    h_{\text{linear}}(t) &= 
    \begin{cases}
        1 - |\frac{t}{N}|, & \text{if } |t| \leq N \\ 
        0, & \text{otherwise} \\
    \end{cases}
    \\
    h_{\text{nearest}}(t) &= 
    \begin{cases}
        1, & \text{if } |t| \leq N \\ 
        0, & \text{otherwise} \\
    \end{cases}
\end{split}
\end{equation}
where $2N + 1$ is the kernel length. The comparison between the ideal and equivalent filter frequency responses is illustrated in Figure~\ref{fig:filter-response}. It can be observed that the frequency responses of these interpolation layers deviate significantly from the ideal rectangular window. In particular, the slow roll-off in the pass-band causes an attenuation of the valid frequency region that should be preserved, represented by the red hatched regions. Meanwhile, the insufficient suppression in the stop-band fails to eliminate the ``mirrored'' aliasing artifacts, indicated by the blue hatched regions. This phenomenon, where the filter fails to preserve the valid frequency region while incompletely removing aliasing artifacts, is known as the ``filter artifact.''

\section{Methodology}
\label{sec:method}

\begin{figure*}[t]
    \centering
    \includegraphics[width=\textwidth]{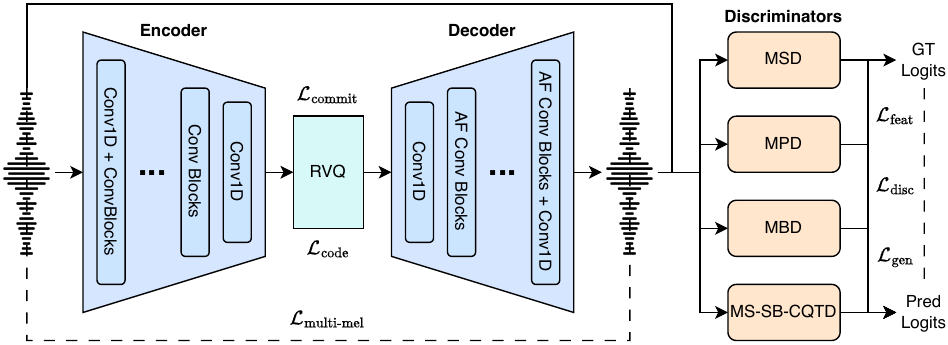}
    \caption{Architecture and training schemes of the proposed models. The Pupu-Codec consists of an encoder, a residual vector quantizer (RVQ) module, a decoder, and four different discriminators. ``AF Conv Blocks'' are obtained by modifying the convolution blocks used in BigVGAN~\cite{bigvgan} and DAC~\cite{dac} with our proposed anti-aliased activation and upsampling modules. Replacing the waveform input, encoder, and RVQ module with a mel-spectrogram as the input gives the Pupu-Vocoder model.}
    \label{fig:model}
\end{figure*}

In this section, we illustrate our idea for obtaining the anti-aliased activation and upsampling modules based on the theoretical analysis in Section~\ref{sec:analysis}, as shown in Figure~\ref{fig:method}.

\subsection{Anti-Aliased Activation Functions}

The architecture of the proposed anti-aliased activation function is shown in Figure~\ref{fig:method}. We use the oversampling~\cite{bigvgan} technique and apply ADAA~\cite{adaa-nonlinear, adaa-origin, adaa-wavedigital, adaa-state, adaa-iir, adaa-rnn} to the activation function. Specifically, we use the SnakeBeta~\cite{snake} activation function to utilize its advantage in modeling the audio's periodic nature, following~\cite{dac, bigvgan}. In particular, the SnakeBeta activation is:
\begin{equation}
    f(x) = x + \frac{\sin^2(\alpha x)}{\beta},
\end{equation}
where $\alpha$ and $\beta$ are learnable parameters. Integrating the equation gives its first-order anti-derivative:
\begin{equation}
    F(x) = \frac{x^2}{2} + \frac{x}{2\beta} - \frac{\sin(2\alpha x)}{4\alpha\beta} + C,
\end{equation}
where $C$ is a constant. Applying ADAA gives:
\begin{equation}
\label{eq:mess}
    y_t = \frac{(\frac{x_t^2}{2} + \frac{x_t}{2\beta} - \frac{\sin(2\alpha x_t)}{4\alpha\beta}) - (\frac{x_{t -1}^2}{2} + \frac{x_{t -1 }}{2\beta} - \frac{\sin(2\alpha x_{t - 1} )}{4\alpha\beta})}{x_t - x_{t - 1}},
\end{equation}
\revision{which can be simplified with the sum-to-product formula, as:
\begin{equation}
\begin{split}
\label{eq:sum-to-product}
    \sin(2\alpha x_t) - \sin(2\alpha x_{t - 1}) = 2\cos(\alpha \Sigma_x)\sin(\alpha \Delta_x).
\end{split}
\end{equation}
where $\Sigma_x = x_t + x_{t - 1}$ and $\Delta_x = x_t - x_{t - 1}$. Substituting Equation~\eqref{eq:sum-to-product} back to the Equation~\eqref{eq:mess} gives:
\begin{equation}
\begin{split}
    y_t = \frac{\Delta_x\Sigma_x} {2\Delta_x} + \frac{\Delta_x}{2\beta\Delta_x} - \frac{\cos(\alpha\Sigma_x)\sin(\alpha\Delta_x)}{2\alpha\beta\Delta_x},
\end{split}
\end{equation}
where each numerator can either eliminate or absorb the denominator, giving the following closed form:
\begin{equation}
\label{eq:adaasnakebeta}
    y_t = \frac{1}{2\beta} + \frac{\Sigma_x}{2} - \frac{\cos(\alpha\Sigma_x)\sinc(\alpha\Delta_x)}{2\beta},
\end{equation}
which yields our ADAA SnakeBeta. As we mentioned in Section~\ref{sec:analysis}, such a function eliminates the denominator entirely, ensuring numerical stability without the need for threshold-based switching in standard ADAA. Meanwhile, it also has bounded outputs for bounded inputs since all $\sin(\cdot)$, $\cos(\cdot)$, $\sinc(\cdot)$, $x_t$, and $x_{t - 1}$ are in $[-1, 1]$. We now investigate the numerical stability of the gradient regarding the ADAA SnakeBeta. Note that for the sinc function, we have:
\begin{equation}
    \sinc'(x) = 
    \begin{cases}
        0, & \text{if } x \rightarrow 0 \\ 
        \frac{\cos(x) - \sinc(x)}{x}, & \text{otherwise } \\
    \end{cases}
\end{equation}
which, when used to take the partial derivative with respect to $x_t$ and $x_{t - 1}$ in Equation~\eqref{eq:adaasnakebeta}, brings the following result:
\begin{equation}
\begin{split}
    \frac{\partial y_t}{\partial x_t} &= \frac{1}{2} + \frac{\alpha}{2\beta}\sin(\alpha\Sigma_x)\sinc(\alpha\Delta_x) \\
    & - \frac{\alpha}{2\beta}\cos(\alpha\Sigma_x)\sinc'(\alpha\Delta_x),
    \\
    \frac{\partial y_t}{\partial x_{t-1}} &= \frac{1}{2} + \frac{\alpha}{2\beta}\sin(\alpha\Sigma_x)\sinc(\alpha\Delta_x) \\ & + \frac{\alpha}{2\beta}\cos(\alpha\Sigma_x)\sinc'(\alpha\Delta_x).
\end{split}
\end{equation}
Applying the auxiliary angle formula gives the following value range of the two partial derivatives:
\begin{equation}
\begin{split}
    \frac{\partial y_t}{\partial x_t} , \frac{\partial y_t}{\partial x_{t-1}} \in \left[\frac{\beta - \alpha}{2\beta}, \frac{\beta + \alpha}{2\beta}\right],
\end{split}
\end{equation}
which mathematically implies a risk of gradient explosion if $\beta \rightarrow 0$ or $\alpha \gg \beta$. This issue can be mitigated through implementation practices and training dynamics. To prevent the $\beta \rightarrow 0$ singularity, we parameterize $\beta$ in the log-domain to enforce a strict lower bound. Meanwhile, the condition $\alpha \gg \beta$ is automatically prevented by the implicit regularization of the loss landscape. Since an excessively large $\alpha$ would introduce a massive amount of meaningless harmonics and an infinitely small $\beta$ would cause the activation values to explode, both scenarios will incur large penalties from the reconstruction and adversarial losses. Thus, combined with standard weight normalization, $\alpha$ and $\beta$ will remain within stable ranges throughout training, ensuring gradient stability in practice.}

\begin{figure}[t]
    \centering
    \begin{subfigure}[b]{0.49\columnwidth}
         \centering
         \includegraphics[width=\columnwidth]{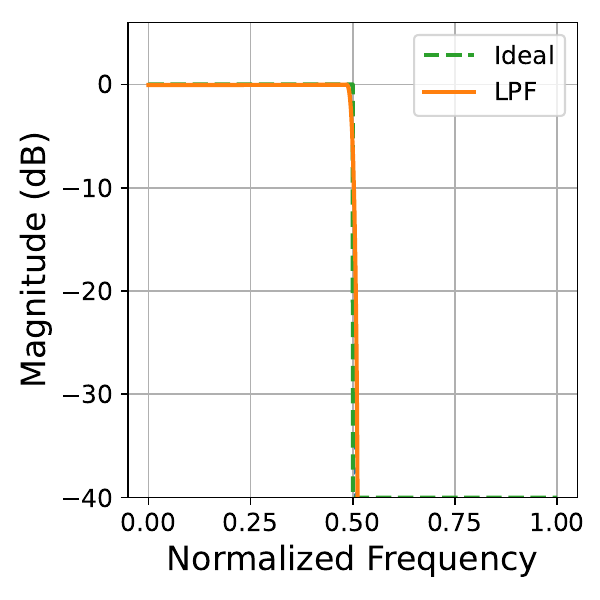}
         \caption{\revision{Low-Pass Filter}}
         \label{fig:lpf-response}
    \end{subfigure}
    \hfill
    \begin{subfigure}[b]{0.49\columnwidth}
         \centering
         \includegraphics[width=\columnwidth]{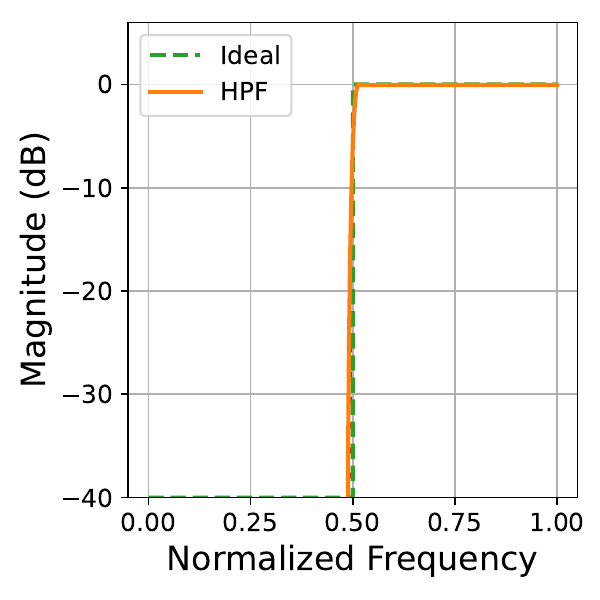}
         \caption{\revision{High-Pass Filter}}
         \label{fig:hpf-response}
    \end{subfigure}
    \caption{\revision{Frequency responses of the filters used in our models. ``LPF'' means low-pass filter and ``HPF'' means high-pass filter. The orange contour and green dashed line represent the actual and ideal frequency responses, respectively.}}
    \vspace{-18pt}
    \label{fig:filter-response-ours}
\end{figure}

\subsection{Anti-Aliased Upsampling Layers}

The architecture of the proposed anti-aliased upsampling layer is shown in Figure~\ref{fig:method}. We replace the ConvTranspose with resampling (zero-interlacing + low-pass filter) to avoid the ``tonal artifact'' and suppress the aliased components. We apply a channel expansion Conv1D layer with a high-pass filter to convert the zero-interlaced $x_0$ to a noise-like deterministic prior, which is used to fill the empty high-frequency region of the upsampled signal to improve the training stability ($x_0$ is the latent representation obtained from the first Conv1D layer in the decoder). The added-up full-band signal is then fed to a channel expansion Conv1D layer to obtain the layer output.  

\revision{Following previous works~\cite{stylegan3, bigvgan, upsample-filter}, we implement the low-pass filter (LPF) by truncating an ideal $\sinc$ function with a Kaiser window, and the filter kernel $h_{\text{LPF}}[n]$ is defined as:
\begin{equation}
    h_{\text{LPF}}[t] =  
    \begin{cases}
        [2f_c \sinc(2f_c(n - N))] \cdot w[n], & \text{if } |t| \leq N \\ 
        0, & \text{otherwise} \\
    \end{cases}
\end{equation}
where $f_c$ is the normalized cut-off frequency and $w[n]$ is the Kaiser window function. Unlike previous works~\cite{stylegan3, bigvgan, upsample-filter}, which used a filter size $n=6$, we use $n=16$ for higher stop-band attenuation. We derive the high-pass filter (HPF) in a complementary manner by subtracting the low-pass filtered signal from the original input signal. As illustrated in Figure~\ref{fig:filter-response-ours}, in contrast to the poor equivalent filter frequency responses of the interpolation layers, our proposed filters exhibit a much closer approximation to the ideal rectangular response.}

\section{Proposed Models}

In this section, we propose Pupu-Vocoder and Pupu-Codec to facilitate audio generation research, as shown in Figure~\ref{fig:model}.

\subsection{Model Architecture}

As illustrated in Figure~\ref{fig:model}, Pupu-Codec consists of an encoder, a residual vector quantizer (RVQ) module, a decoder, and four different discriminators. The encoder includes an initial Conv1D layer, five CNN-based Resblocks, and a final Conv1D layer. The RVQ module is adapted from DAC~\cite{dac}. The decoder includes an initial Conv1D layer, five ``AF Conv Blocks'', and a final Conv1D layer. The ``AF Conv Blocks'' are obtained by modifying the convolution blocks used in BigVGAN~\cite{bigvgan} and DAC~\cite{dac} with our proposed anti-aliased activation and upsampling modules. Following~\cite{cqtjournal}, we mix both time-domain and TFR-based discriminators to obtain a better synthesis quality, which includes: Multi-Period Discriminator (MPD), Multi-Scale Discriminator (MSD)~\cite{HiFiGAN}, Multi-Band Discriminator (MBD)~\cite{dac}, and Multi-Scale Sub-Band CQT Discriminator (MS-SB-CQTD)~\cite{cqt}. Replacing the waveform input, encoder, and RVQ module with a mel-spectrogram as the input gives the Pupu-Vocoder model.

\subsection{Training Losses}

We adapted the training scheme from DAC~\cite{dac} to train the Pupu-Codec model, which is illustrated as follows:
\begin{equation}
\begin{split}
    &\mathcal{L}_\text{generator} = 15\mathcal{L}_\text{multi-mel}(x, \hat{x}) + 0.25\mathcal{L}_\text{commit}(z_q, z_e) \\
    & +\mathcal{L}_\text{code}(z_e, z_q) + \sum_{m=1}^{M}[\mathcal{L}_\text{adv}(G; D_{m}) + 2\mathcal{L}_\text{feat}(G; D_m)];
    \\
    &\mathcal{L}_\text{discriminator} = \sum_{m=1}^{M}\mathcal{L}_\text{adv}(D_{m}; G); 
\end{split}
\end{equation}
where $x$ and $\hat{x}$ are the ground truth and predicted waveform, $z_e$ is the latent representation from encoder, $z_q$ is the quantized $z_e$ reconstructed from discrete tokens, $\mathcal{L}_\text{code}$, $\mathcal{L}_\text{commit}$, and $\mathcal{L}_\text{multi-mel}$ are the codebook, commitment, and multi-scale mel-spectrogram losses, $D_m$ is the $m_\text{th}$ discriminator, $G$ is the Pupu-Codec model, and $\mathcal{L}_\text{adv}$ and $\mathcal{L}_\text{feat}$ are the adversarial losses and feature matching loss. Removing the codebook and commitment losses gives the training goal of Pupu-Vocoder. 

\section{Experiments}

\subsection{Experiment Setup}

\subsubsection{Datasets}

As suggested by previous works~\cite{emilia, foleycrafter, sslbuscomp, neurodyne, emilia-journal}, we use large-scale data mixtures from different domains to train and evaluate the models for better distinguishability. Specifically, for speech, we use DAPS~\cite{daps}, HQ-TTS~\cite{hqtts}, AIShell 3~\cite{aishell3}, HiFi-TTS~\cite{hifitts}, HUI-TTS~\cite{huitts}, VCTK~\cite{vctk}, Bible-TTS~\cite{bibletts}, EARS~\cite{ears}, and Mana-TTS~\cite{manatts} for training, resulting in 1661.4 hours of multilingual speech; for singing voice, we use NUS-48E~\cite{NUS48E}, Opera~\cite{opera}, VocalSet~\cite{VocalSet}, JSUTSong~\cite{JSUTSong}, JaCRC~\cite{jacrc}, PJS~\cite{pjs}, CSD~\cite{csd}, JVS-Music~\cite{JVS-MuSiC}, KiSing~\cite{kising}, OpenSinger~\cite{OpenSinger}, NHSS~\cite{NHSS}, PopCS~\cite{diffsinger}, PopBuTFy~\cite{popbutfy}, Opencpop~\cite{Opencpop}, M4Singer~\cite{M4Singer}, SingStyle111~\cite{SingStyle}, GOAT~\cite{goat}, ACESinger~\cite{ACESinger}, SingNet-SP~\cite{singnet}, and an internal dataset for training, resulting in 885.2 hours of multilingual and multi-style singing voice; for music, we use GoodSounds~\cite{goodsounds}, MedleyDB~\cite{medleydb}, MUSDB18~\cite{musdbhq}, Slakh2100~\cite{slakh}, Surge Synth~\cite{surge}, Arturia Synth~\cite{surge}, DX7 Synth~\cite{surge}, and MoisesDB~\cite{moisesdb} for training, resulting in 2343.1 hours of multi-style music tracks; for audio, we use Audioset-Strong~\cite{audiosetstrong}, BBC Sound Effects~\footnote{\url{https://sound-effects.bbcrewind.co.uk/}}, and FreeSound~\footnote{\url{https://freesound.org/}} for training, resulting in 1811.0 hours audio events. To evaluate the performance of models in different domains, we split the evaluation sets into academic and industrial settings. The academic setting is formed by academic datasets, providing a diverse yet general evaluation, while the industrial setting is formed by a small amount of industrial data, offering a high-quality and professional evaluation. Specifically, for speech, we utilize English, Chinese, Japanese, Korean, French, and German sets from Common Voice~\cite{commonvoice} as the academic setting. We manually collected voice actor databases from Hitsugi\footnote{\url{https://booth.pm/ja/items/3382115}}, ZunzunProject\footnote{\url{https://zunko.jp/}}, Voice Seven\footnote{\label{footnote:voiceseven}\url{https://voiceseven.com/}}, Amitaro\footnote{\url{http://amitaro.net/}}, Narakuyui\footnote{\url{https://narakuyui.fanbox.cc/posts/7082575}}, and Matsukane\footnote{\url{https://x.com/mochi_jin_voice}} as the industrial setting; for singing voice, we use GTSinger~\cite{gtsinger} as the academic setting. We collected Vocaloid databases from Kiritan~\cite{kiritan}, Namine Ritsu\footnote{\url{https://www.canon-voice.com/}}, Voice Seven\footnoteref{footnote:voiceseven}, Oniku Kurumi\footnote{\url{https://onikuru.info/}}, Ofutonp\footnote{\url{https://sites.google.com/view/oftn-utagoedb/}}, Yuuri Natsume\footnote{\url{https://ksdcm1ng.wixsite.com/njksofficial}}, and Amaboshi Cipher\footnote{\url{https://parapluie2c56m.wixsite.com/mysite}} for industrial setting; for music, we use the Cambridge Mixing Secret\footnote{\url{https://www.cambridge-mt.com/ms3/mtk/}} database as the academic setting. We utilize an internal music production sample pack database for the industrial setting. For audio, we employ the UrbanSound8K~\cite{urbansound}, ECE50~\cite{ece50}, and MACS~\cite{macs} datasets for the academic setting. We utilize an internal sound design sample pack database for industrial settings. Each academic or industrial evaluation set is constructed of 1000 samples that are evenly and randomly extracted from each data source.

We also constructed a test signal benchmark to demonstrate the effectiveness of the anti-aliased modules, following~\cite{asr}. In particular, we use three test signal types: sine, sawtooth, and triangle waves. We use Serum\footnote{\url{https://xferrecords.com/products/serum-2}} to generate the test signals to utilize its outstanding anti-aliasing ability. We use Reaper\footnote{\url{https://www.reaper.fm/}} and Reascript\footnote{\url{https://www.reaper.fm/sdk/reascript/reascript.php}} for automatic command-line audio generation. For each signal, we generate 10-second MIDI notes from C4 (261.63\,Hz) to B7 (3951.04\,Hz), which will then be symmetrically trimmed at the beginning and end to eliminate the clicks caused by the attack and release phases in the attack-decay-sustain-release (ADSR) generation process, resulting in 48 5-second segments.
\revision{It is worth noting that we rely on these synthetic test signals because they are stationary and periodic, which makes it easy to separate and track harmonic components from aliasing artifacts, thereby yielding effective AHR values. On the other hand, estimating AHR from real-world signals would require harmonic tracking and special treatment of non-harmonic signal content, which is not trivial and would likely yield unreliable results.}


\subsubsection{Preprocessing}

We process the training and evaluation datasets to 44.1\,kHz mono WAV files. For extracting the mel-spectrograms, we use an FFT size of 2048, a hop size of 512, a window length of 2048, and 128 mel filters, which are further normalized in log-scale with values $\leq$ 1e-5 clipped to 0.

\subsubsection{Configurations}

The Pupu-Codec is modified from DAC~\cite{dac}. Based on the official repository\footnote{\label{footnote:dac}\url{https://github.com/descriptinc/descript-audio-codec}}, we change the encoder channel dimension to 32 for the small model and to 48 for the large model. We change the encoder and decoder ratios to [2, 2, 2, 8, 8], and encoder and decoder kernel sizes to [4, 4, 4, 16, 16], and leave the RVQ module unmodified to have a frame rate of 86\,Hz and a maximum bitrate of 8\,kbps. The Pupu-Vocoder is modified from BigVGAN~\cite{bigvgan}. Based on the official repository\footnote{\label{footnote:bigvgan}\url{https://github.com/NVIDIA/BigVGAN}}, we modify the upsampling ratios to [8, 8, 2, 2, 2], upsampling kernel sizes to [16, 16, 4, 4, 4], and leave the initial channel size unmodified. For the anti-aliased activation and upsampling modules, we apply ADAA with an oversampling factor of 2 for the activation function. For the upsampling layer, we use a kernel size of 7, a stride of 1, and a padding of 1 for the noise convolution, and a kernel size of 1, a stride of 1, and a padding of 1 for the channel expansion convolution. For the discriminators, we use periods of [2, 3, 5, 7, 11, 17, 23, 37] for the MPD. We compute 10 octaves and change the hop sizes to [1024, 512, 512] for the MS-SB-CQTD, while leaving the MSD and MBD unmodified.

\subsubsection{Baselines}

We use various baselines to illustrate the effectiveness of our proposed models. We use HiFi-GAN~\cite{HiFiGAN} and BigVGAN~\cite{bigvgan} as neural vocoder baselines and use Encodec~\cite{encodec}, DAC~\cite{dac}, and BigCodec~\cite{bigcodec} as neural codec baselines. We additionally use Vocos~\cite{vocos} as the aliasing-free TF-domain referential system. We maintain all the codec systems at the same frame rate and maximum bitrate for a fair comparison by adjusting their encoder and decoder ratios.

The detailed model configurations are illustrated as follows:

\begin{itemize}
    \item \textbf{HiFi-GAN} - We implement the HiFi-GAN model using\footnote{\url{https://github.com/jik876/hifi-gan}} and modify the upsampling ratios to [8, 8, 2, 2, 2], upsampling kernel sizes to [16, 16, 4, 4, 4], and the initial channel size to 512 based on the V1 model.
    \item \textbf{BigVGAN} - We implement the BigVGAN using\footref{footnote:bigvgan}. We modify the upsampling ratios to [8, 8, 2, 2, 2] and the upsampling kernel sizes to [16, 16, 4, 4, 4] for the small model, while keeping the large model unchanged.
    \item \textbf{Encodec} - We implement the Encodec model using an open-source reproduction\footnote{\url{https://github.com/ZhikangNiu/encodec-pytorch}} and modify the encoder and decoder ratios to [2, 2, 2, 8, 8], encoder and decoder kernel sizes to [4, 4, 4, 16, 16], and target bandwidths to be [1.78\,kbps, 2.67\,kbps, 5.33\,kbps, 8\,kbps].
    \item \textbf{DAC} - We implement the DAC using\footref{footnote:dac} and modify the encoder and decoder ratios to [2, 2, 2, 8, 8], and encoder and decoder kernel sizes to [4, 4, 4, 16, 16].
    \item \textbf{BigCodec} - We implement the BigCodec using\footnote{\url{https://github.com/Aria-K-Alethia/BigCodec}} and modify the encoder and decoder ratios to [2, 2, 2, 2, 4, 8], encoder and decoder kernel sizes to [4, 4, 4, 4, 8, 16], codebook size to 1024, and codebook number to 9.
    \item \textbf{Vocos} - We implement the Vocos model using\footnote{\url{https://github.com/gemelo-ai/vocos}} and modify the upsampling ratios to [8, 8, 2, 2, 2], and upsampling kernel sizes to [16, 16, 4, 4, 4].
\end{itemize}

\rerevision{Note that all baseline models selected and re-implemented for comparison are without the fundamental frequency (F0) conditioning}, since general audio and music lack an explicit F0, and omitting it across all domains ensures consistency and prevents the unfair performance bias that conditioned models would otherwise exhibit on speech and singing voice.

\begin{figure*}[t]
    \centering
    \begin{subfigure}[b]{0.19\textwidth}
         \centering
         \includegraphics[width=\textwidth]{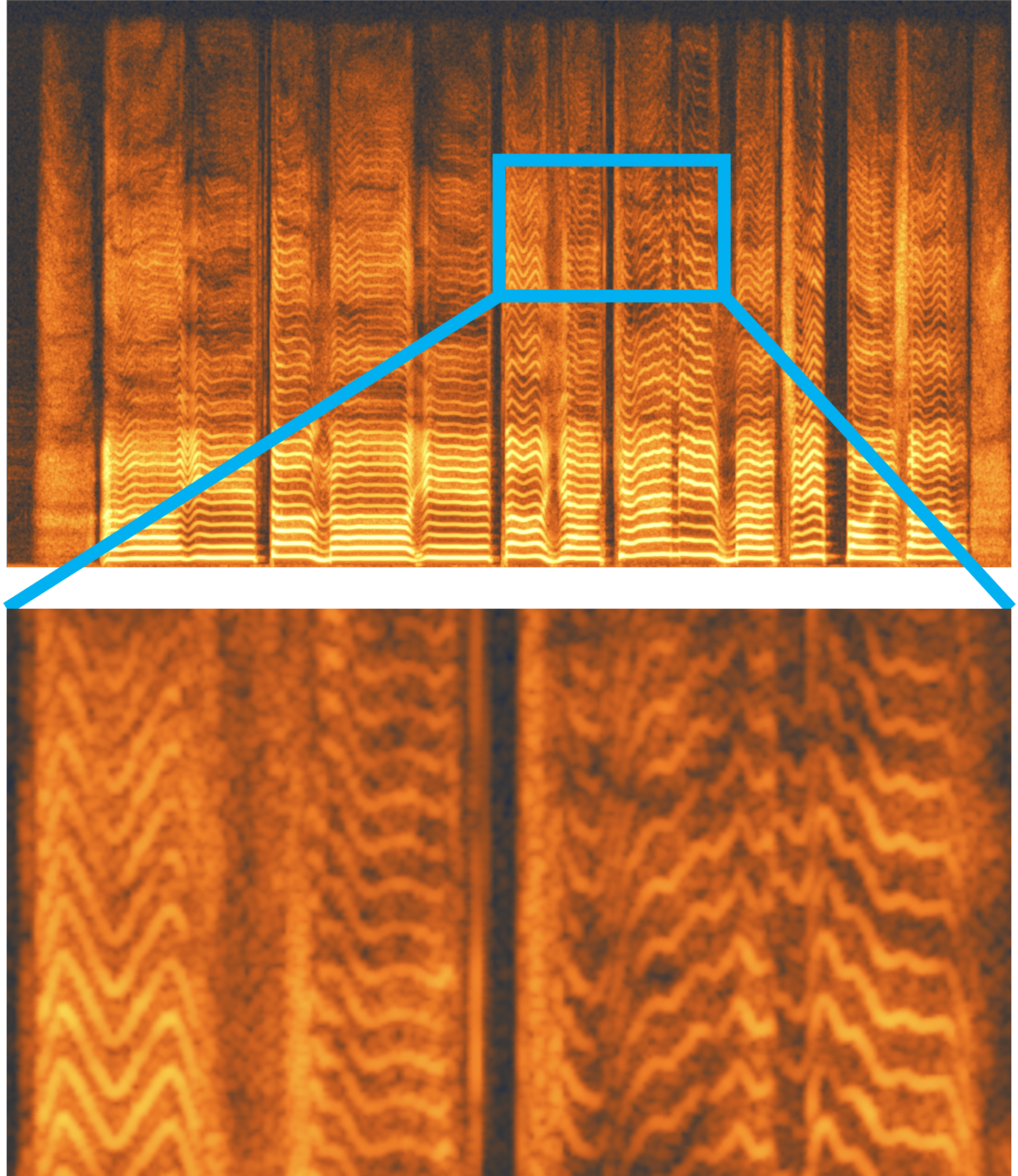}
         \caption{Ground Truth}
         \label{fig:gt-singing}
    \end{subfigure}
    \hfill
    \begin{subfigure}[b]{0.19\textwidth}
         \centering
         \includegraphics[width=\textwidth]{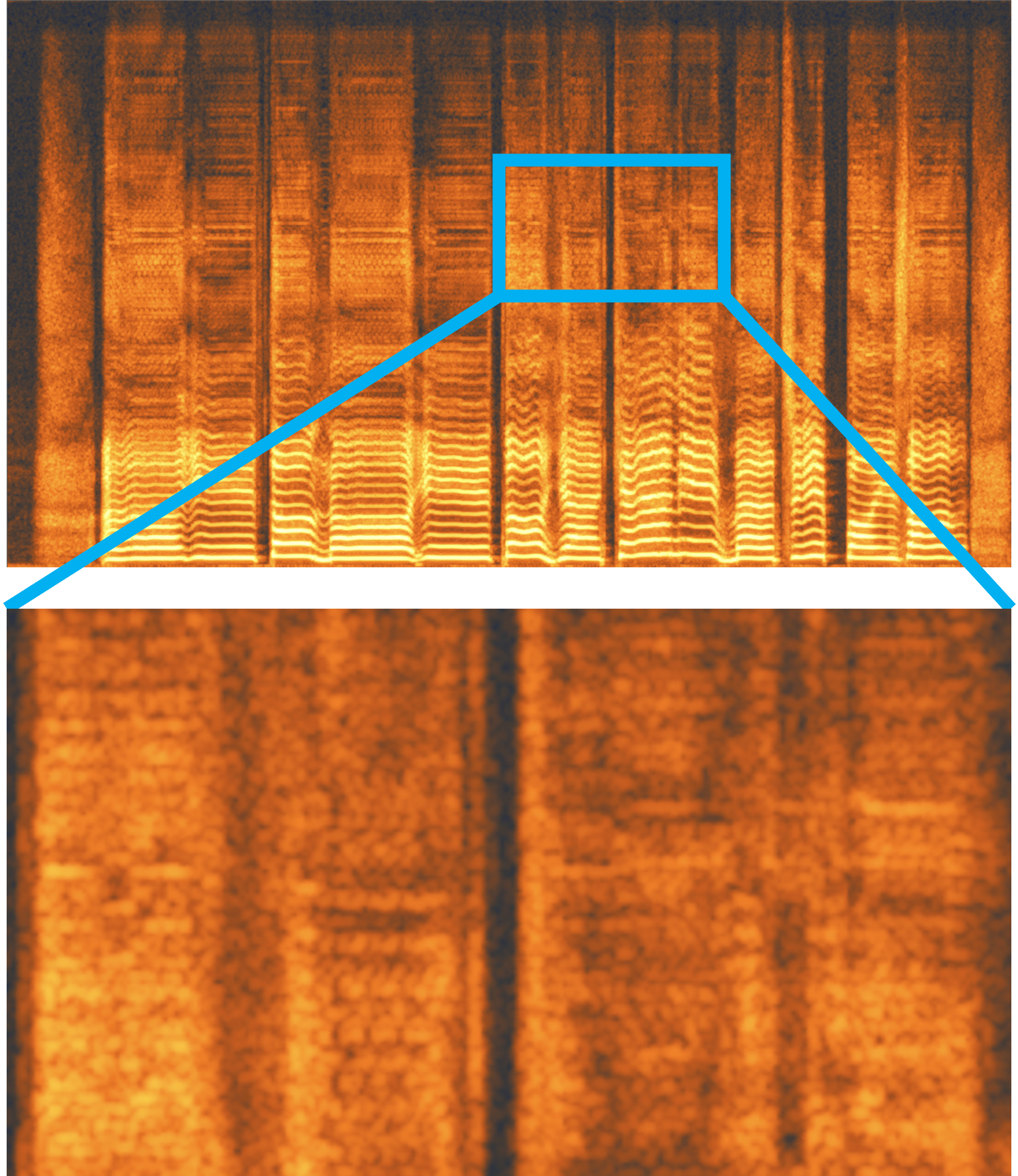}
         \caption{Vocos}
         \label{fig:vocos-singing}
    \end{subfigure}
    \hfill
    \begin{subfigure}[b]{0.19\textwidth}
         \centering
         \includegraphics[width=\textwidth]{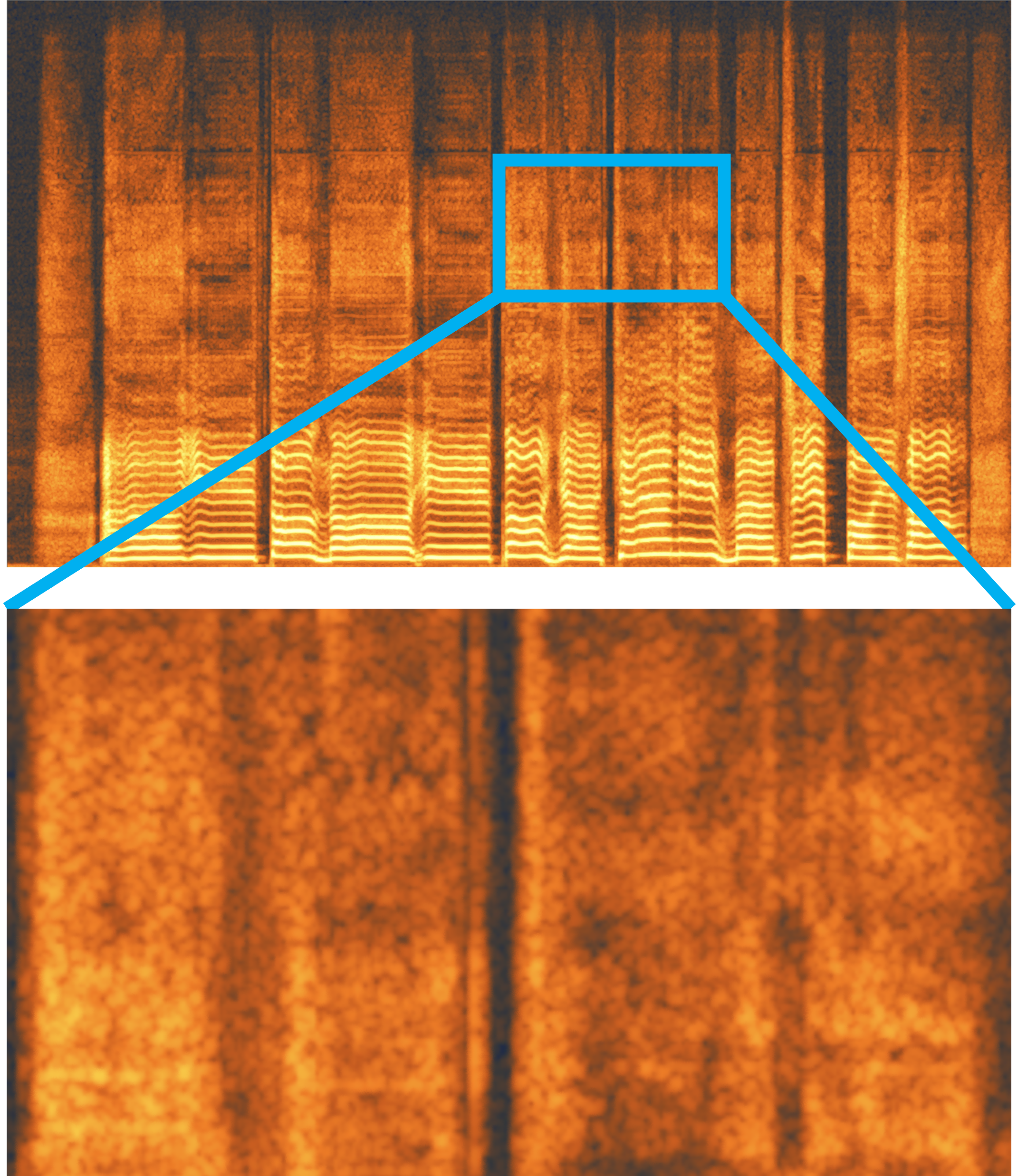}
         \caption{BigVGAN$_\text{small}$}
         \label{fig:bigvgan_small-singing}
    \end{subfigure}
    \hfill
    \begin{subfigure}[b]{0.19\textwidth}
         \centering
         \includegraphics[width=\textwidth]{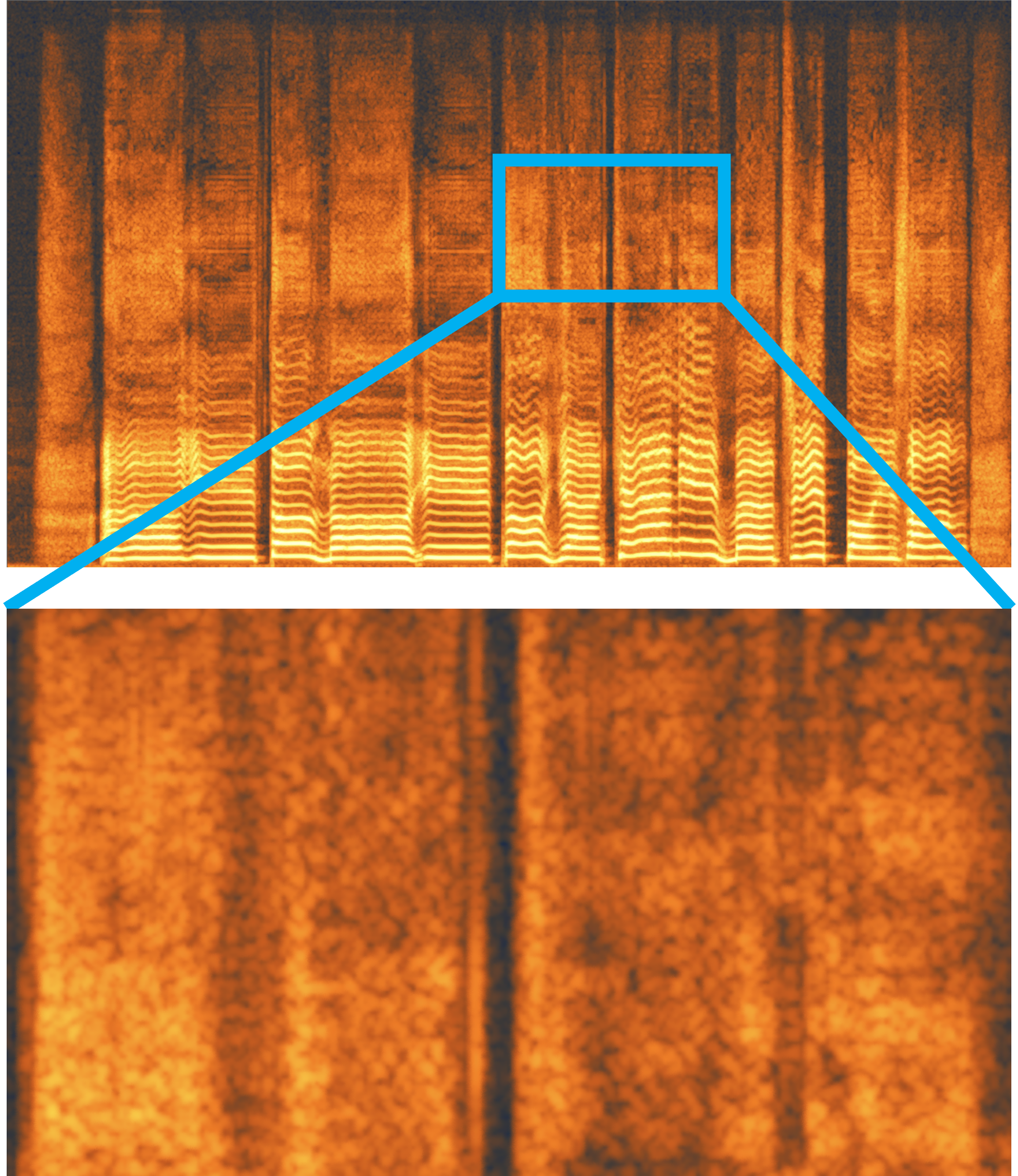}
         \caption{BigVGAN$_\text{large}$}
         \label{fig:bigvgan_large-singing}
    \end{subfigure}
    \hfill
    \begin{subfigure}[b]{0.19\textwidth}
         \centering
         \includegraphics[width=\textwidth]{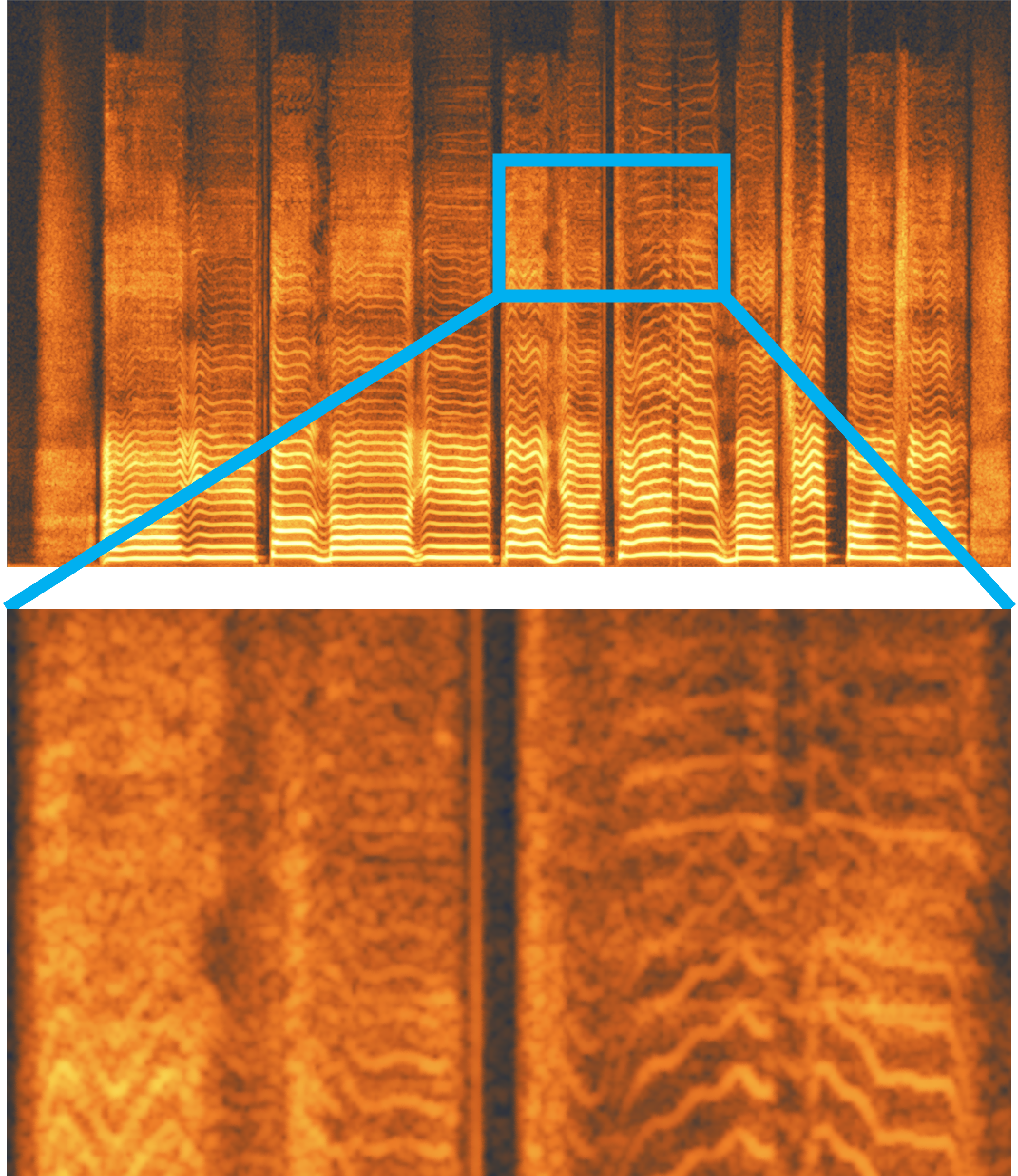}
         \caption{DAC}
         \label{fig:dac-singing}
    \end{subfigure}
    \\
    \begin{subfigure}[b]{0.19\textwidth}
         \centering
         \includegraphics[width=\textwidth]{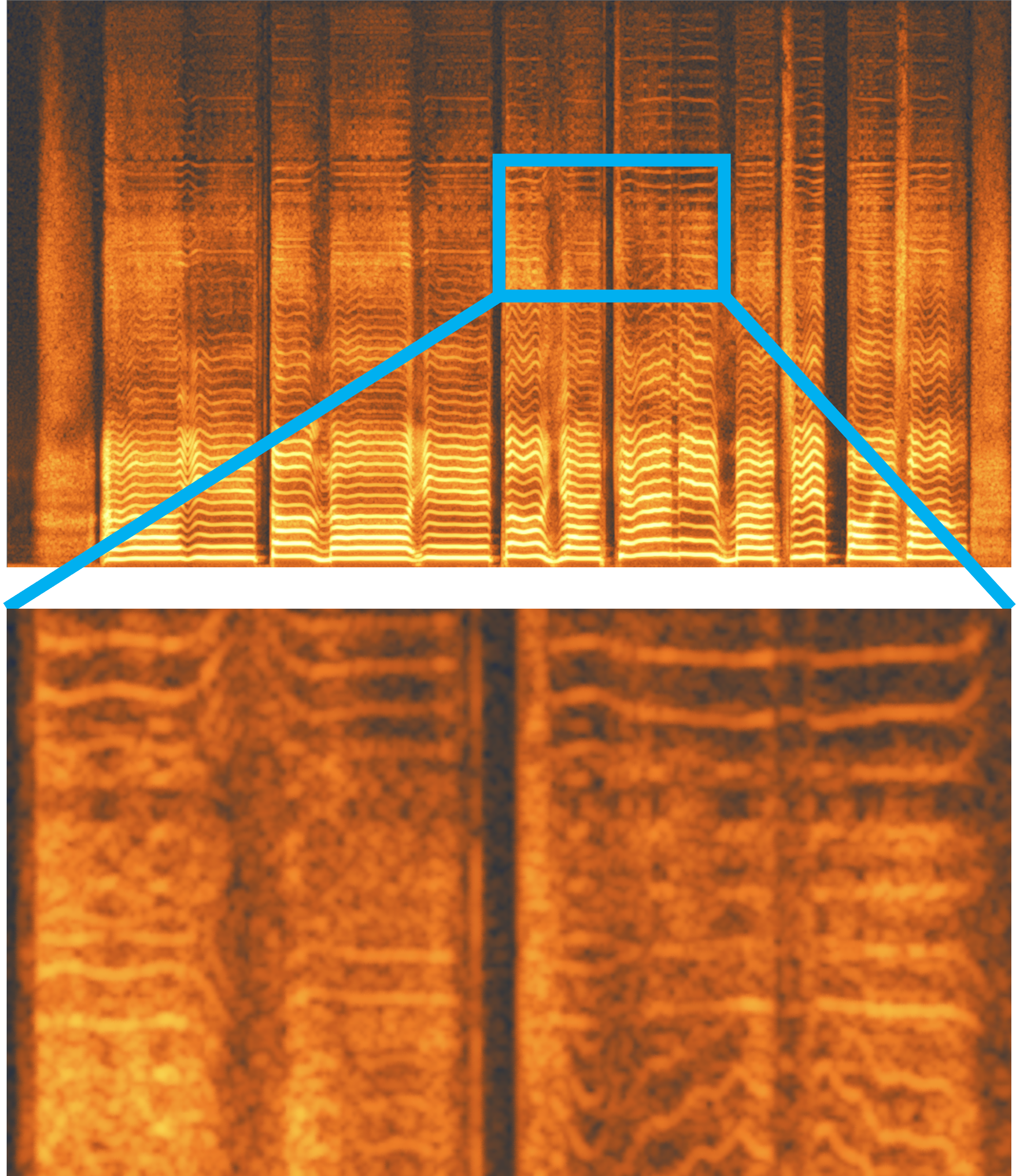}
         \caption{BigCodec}
         \label{fig:bigcodec-singing}
    \end{subfigure}
    \hfill
    \begin{subfigure}[b]{0.19\textwidth}
         \centering
         \includegraphics[width=\textwidth]{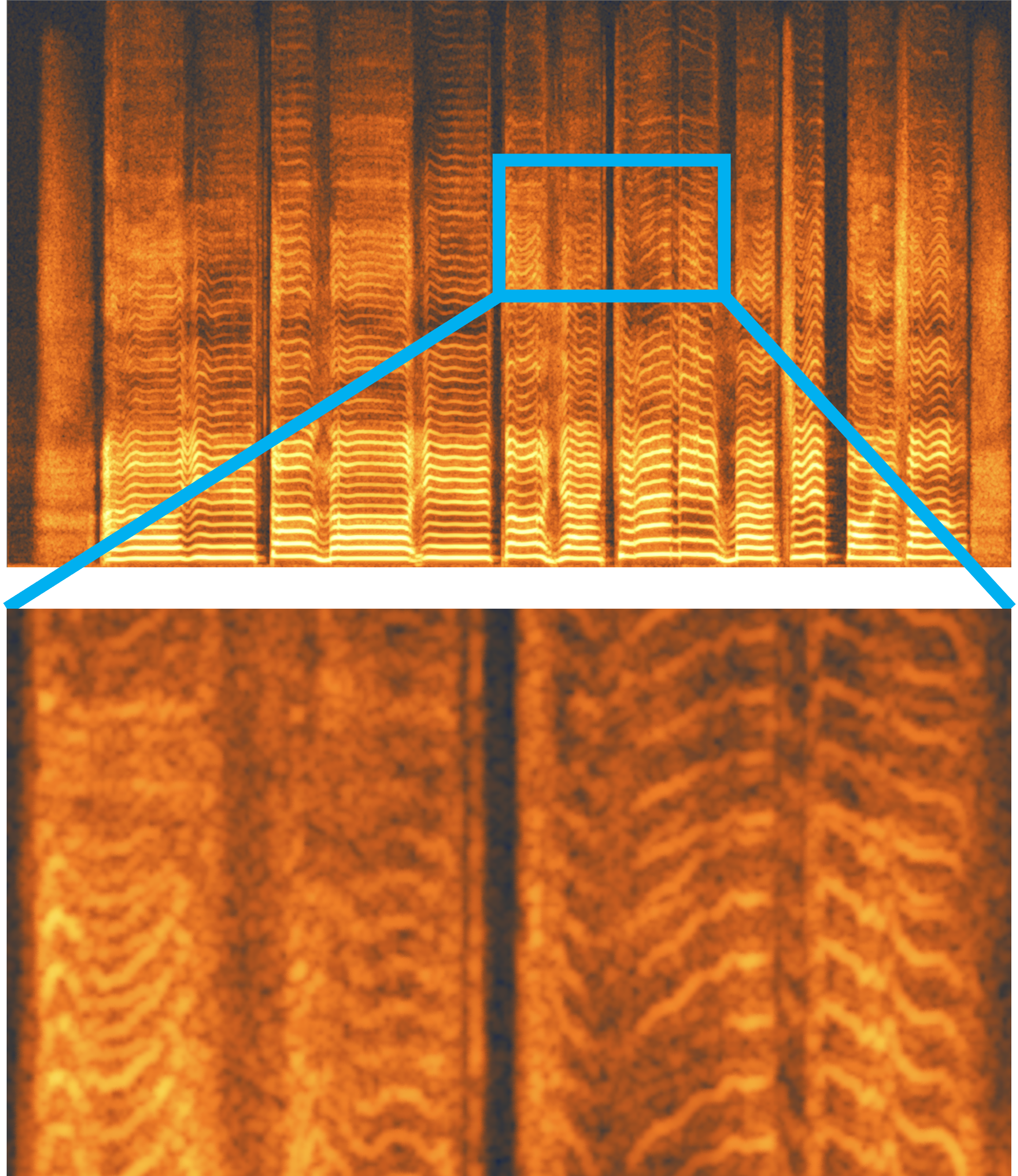}
         \caption{Pupu-Vocoder$_\text{small}$}
         \label{fig:pupuvocoder_small-singing}
    \end{subfigure}
    \hfill
    \begin{subfigure}[b]{0.19\textwidth}
         \centering
         \includegraphics[width=\textwidth]{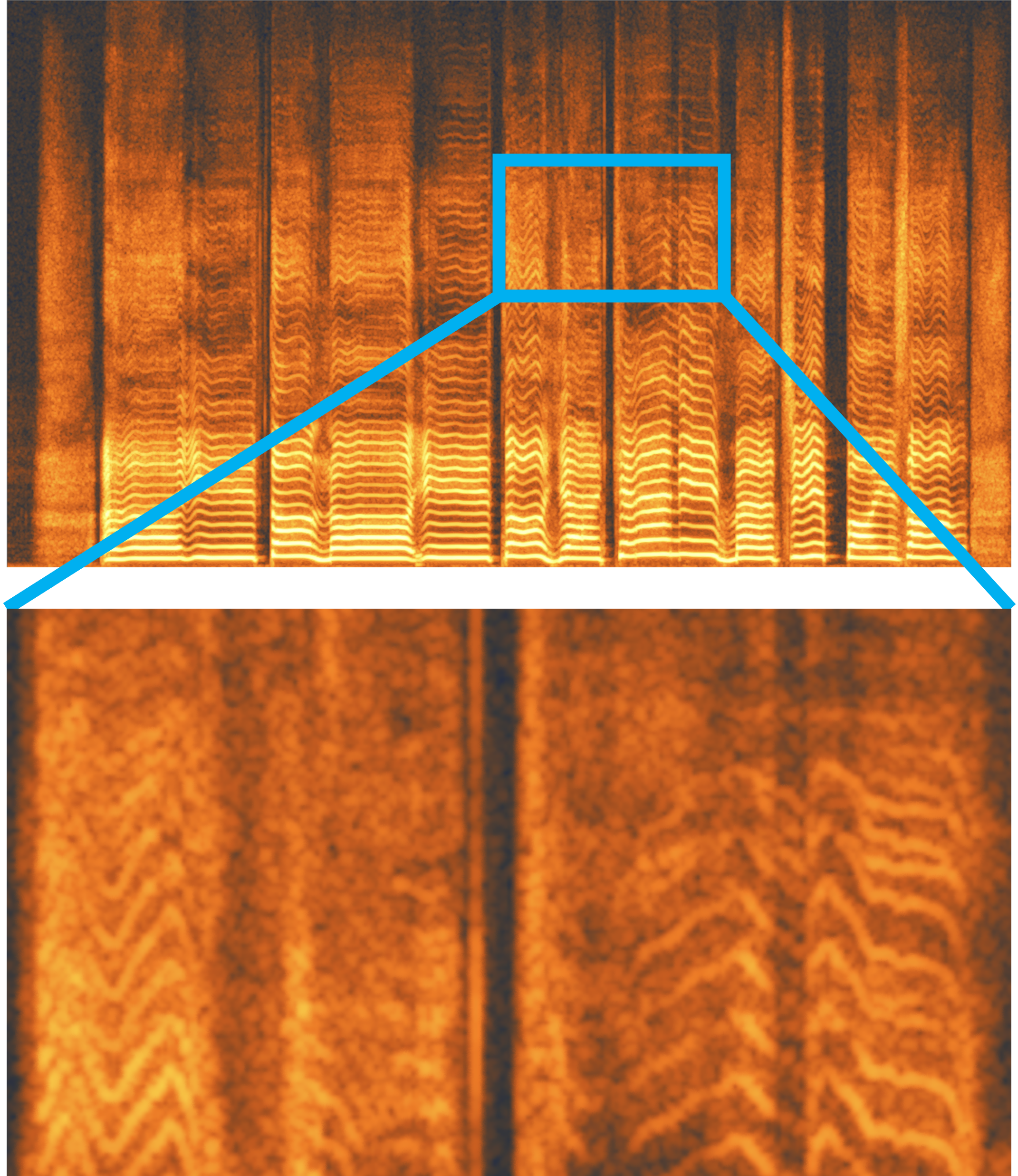}
         \caption{Pupu-Vocoder$_\text{large}$}
         \label{fig:pupuvocoder_large-singing}
    \end{subfigure}
    \hfill
    \begin{subfigure}[b]{0.19\textwidth}
         \centering
         \includegraphics[width=\textwidth]{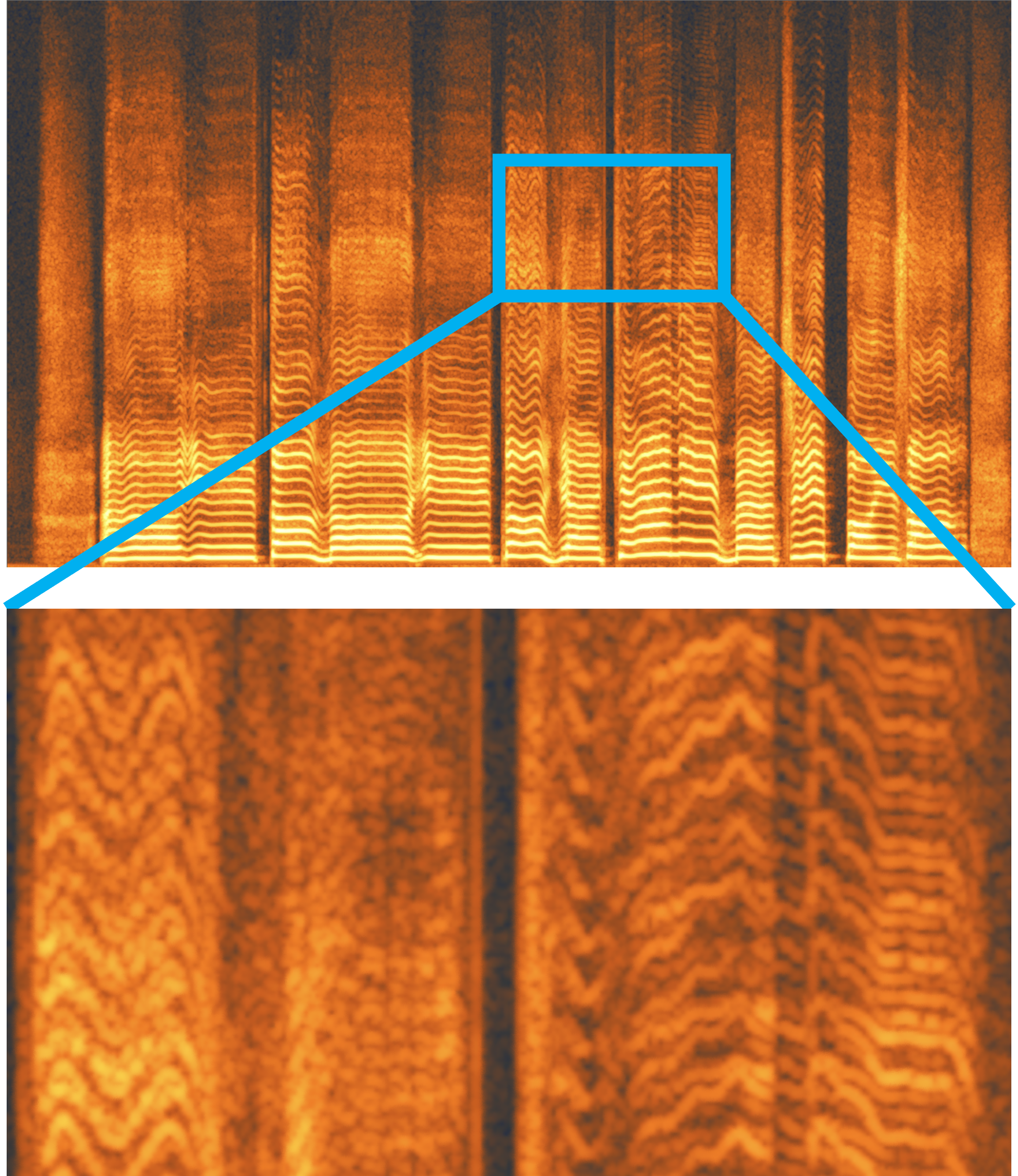}
         \caption{Pupu-Codec$_\text{small}$}
         \label{fig:pupucodec_small-singing}
    \end{subfigure}
    \hfill
    \begin{subfigure}[b]{0.19\textwidth}
         \centering
         \includegraphics[width=\textwidth]{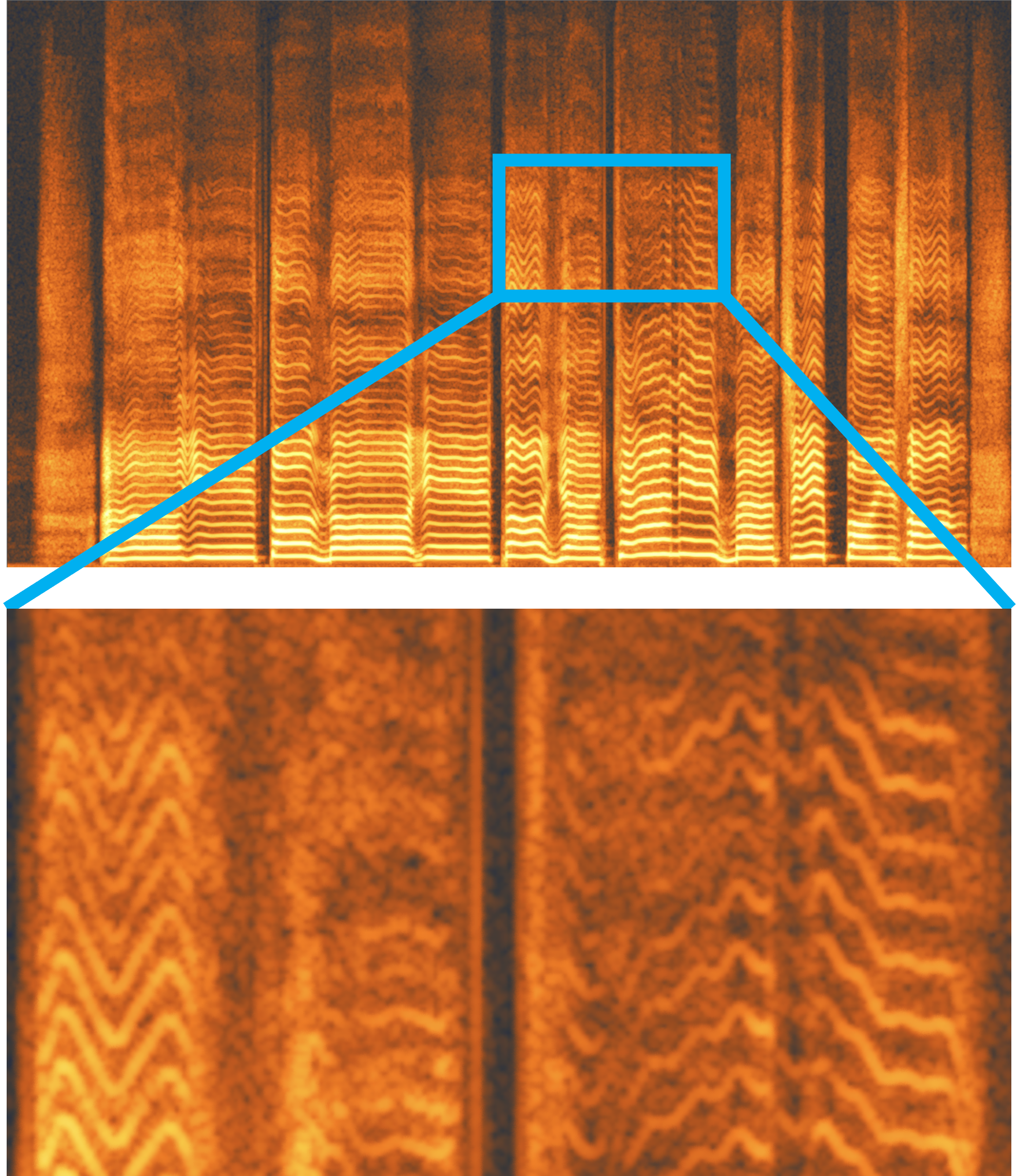}
         \caption{Pupu-Codec$_\text{large}$}
         \label{fig:pupucodec_large-singing}
    \end{subfigure}
    \caption{Spectrogram visualization with a zoomed-in view of high-frequency harmonic components (around 16\,kHz) regarding a representative singing voice example copy-synthesized by different neural vocoder and codec models.}
    \vspace{-14pt}
    \label{fig:singing-voice-case-study}
\end{figure*}

\subsubsection{Training}

All the models are trained using the AdamW optimizer with $\beta _ {1} = 0.8$ and $\beta _ {2} = 0.99$, a learning rate of 1e-4, and an exponential decay scheduler with a factor $\gamma = 0.999996$. All the experiments are conducted on 8 H200 GPUs with the maximum available batch size for 1M steps.

\subsection{Evaluation Metrics}

\subsubsection{Objective Evaluation}

We use the Amphion~\cite{amphion} toolkit for objective evaluation. For the test signal benchmark, we use the aliasing-to-harmonic ratio (AHR) as proposed in~\cite{asr}, with a definition that matches the aliasing-to-signal ratio in the original paper, but avoids an abbreviation clash with Automatic Speech Recognition (ASR). For the speech and singing voice, we use the predictive mean opinion score (MOS-Pred), F0 root mean square error (F0RMSE), and F0 periodicity (Periodicity) following~\cite{contentsvc, SVCC}. \rerevision{We use the TorchFCPE~\cite{torchfcpe} toolkit as our F0 predictor, and we compute all F0-related metrics exclusively in the voiced region and report their value in cents.} We use the Sheet~\cite{sheet} toolkit to predict the MOS value. For music and audio, we use the Fr\'{e}chet audio distance (FAD)~\cite{fad, fad-music}, multi-scale STFT distance (M-STFT), and ViSQOL~\cite{visqol} following DAC~\cite{dac}. We use MERT~\cite{mert} as the feature extractor for the FAD model. \revision{To evaluate inference efficiency, we compute the real-time factor (RTF) using an NVIDIA H200 GPU and an Intel Xeon Platinum 8581C CPU. Note that the Sheet~\cite{sheet} MOS predictor is limited to 16\,kHz, as we use it as a superior alternative to rigid metrics such as Mel-Cepstral Distortion (MCD) for evaluating the mid- to low-frequency bands. The actual 44.1\,kHz full-band synthesis quality is evaluated through our subjective MUSHRA tests. We do not use the recent 44.1\,kHz-compatible MOS predictors~\cite{audiomos}, as most remain closed-source and their performance is limited by a lack of singing voice training data.}

\begin{figure*}[t]
    \captionsetup[subfigure]{skip=3pt}
    \centering
    \begin{subfigure}[b]{0.49\columnwidth}
         \centering
         \includegraphics[width=\columnwidth, trim={0 0.4cm 0 0},clip]{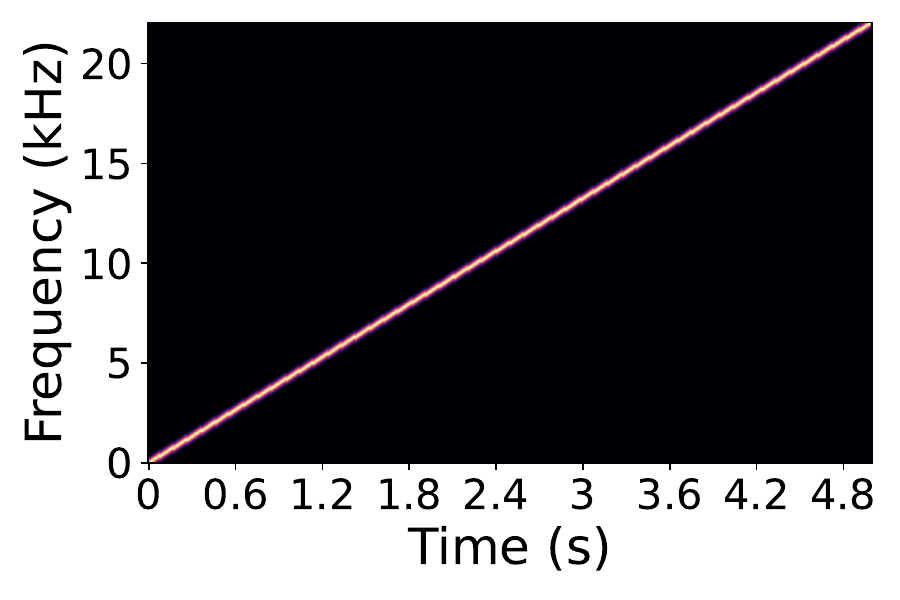}
         \caption{No activation}
         \label{fig:sinesweep}
    \end{subfigure}
    \begin{subfigure}[b]{0.49\columnwidth}
         \centering
         \includegraphics[width=\columnwidth, trim={0 0.4cm 0 0},clip]{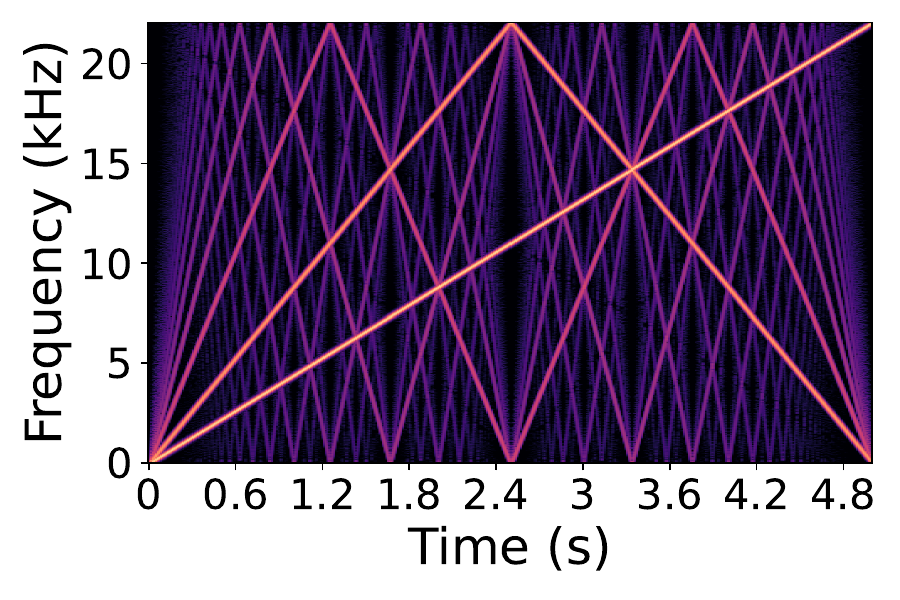}
         \caption{LeakyReLU}
         \label{fig:leakyrelu}
    \end{subfigure}
    \begin{subfigure}[b]{0.49\columnwidth}
         \centering
         \includegraphics[width=\columnwidth, trim={0 0.4cm 0 0},clip]{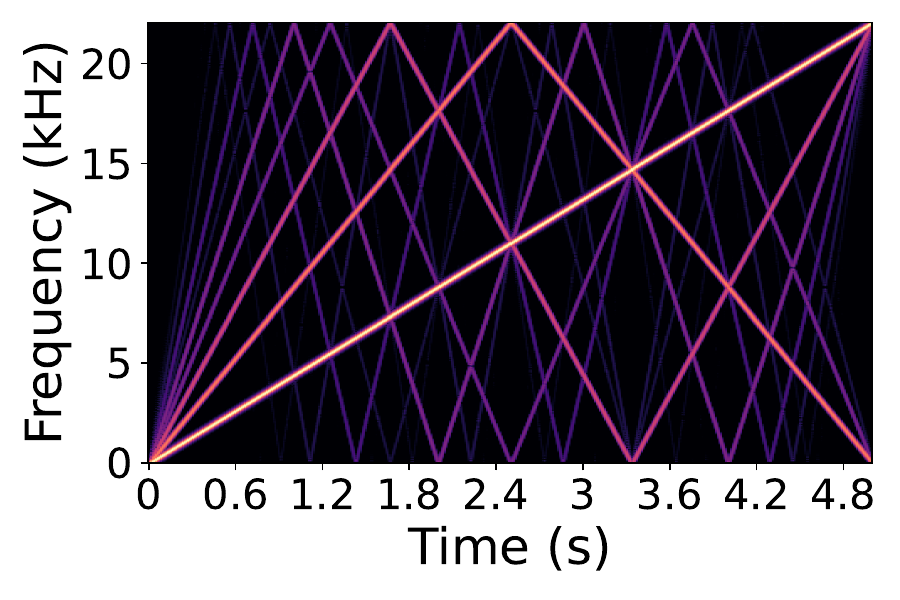}
         \caption{ELU}
         \label{fig:elu}
    \end{subfigure}
    \begin{subfigure}[b]{0.49\columnwidth}
         \centering
         \includegraphics[width=\columnwidth, trim={0 0.4cm 0 0},clip]{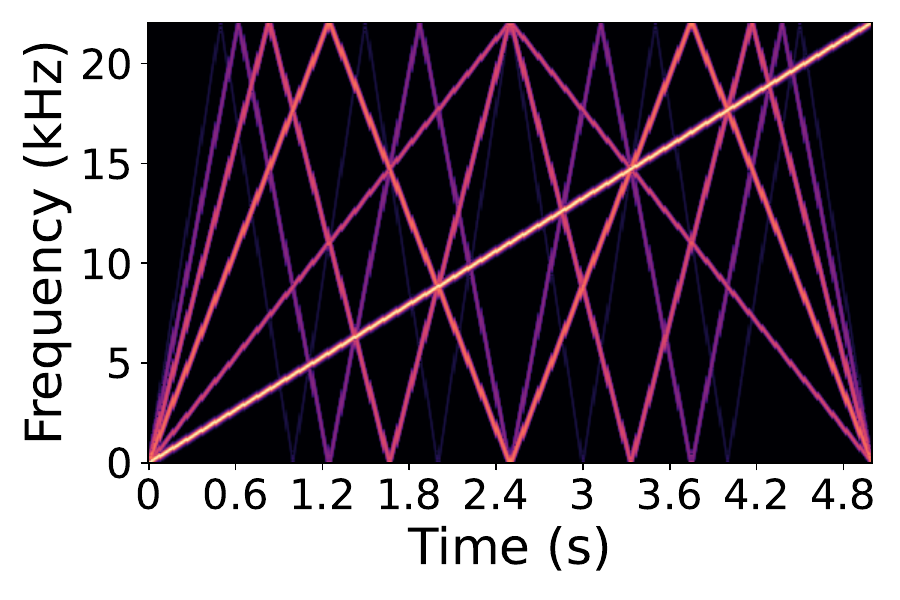}
         \caption{SnakeBeta}
         \label{fig:snakebeta}
    \end{subfigure}
    \\
    \begin{subfigure}[b]{0.49\columnwidth}
         \centering
         \includegraphics[width=\columnwidth, trim={0 0.4cm 0 0},clip]{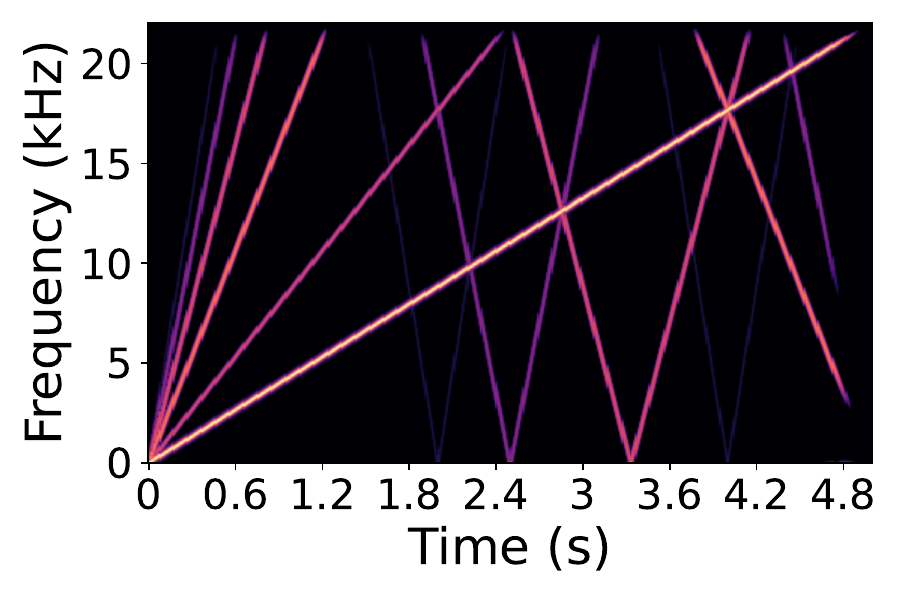}
         \caption{SnakeBeta ($O=2$)}
         \label{fig:snakebeta_2}
    \end{subfigure}
    \begin{subfigure}[b]{0.49\columnwidth}
         \centering
         \includegraphics[width=\columnwidth, trim={0 0.4cm 0 0},clip]{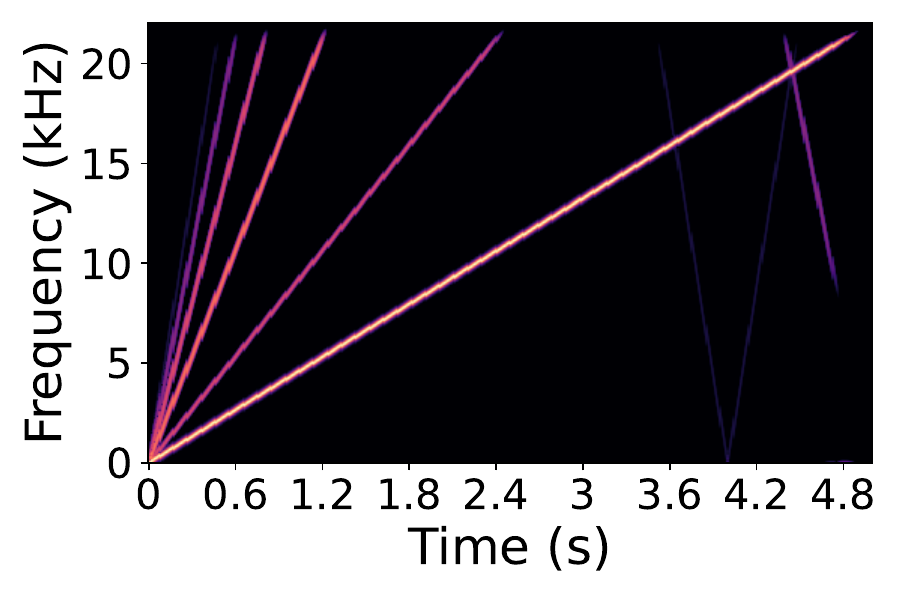}
         \caption{SnakeBeta ($O=4$)}
         \label{fig:snakebeta_4}
    \end{subfigure}
    \begin{subfigure}[b]{0.49\columnwidth}
         \centering
         \includegraphics[width=\columnwidth, trim={0 0.4cm 0 0},clip]{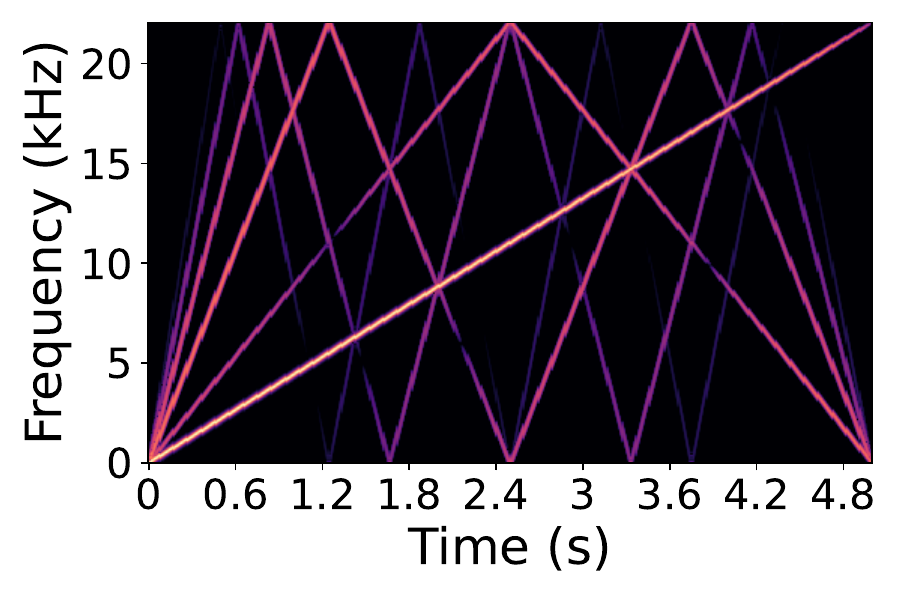}
         \caption{ADAA SnakeBeta}
         \label{fig:adaasnakebeta}
    \end{subfigure}
    \begin{subfigure}[b]{0.49\columnwidth}
         \centering
         \includegraphics[width=\columnwidth, trim={0 0.4cm 0 0},clip]{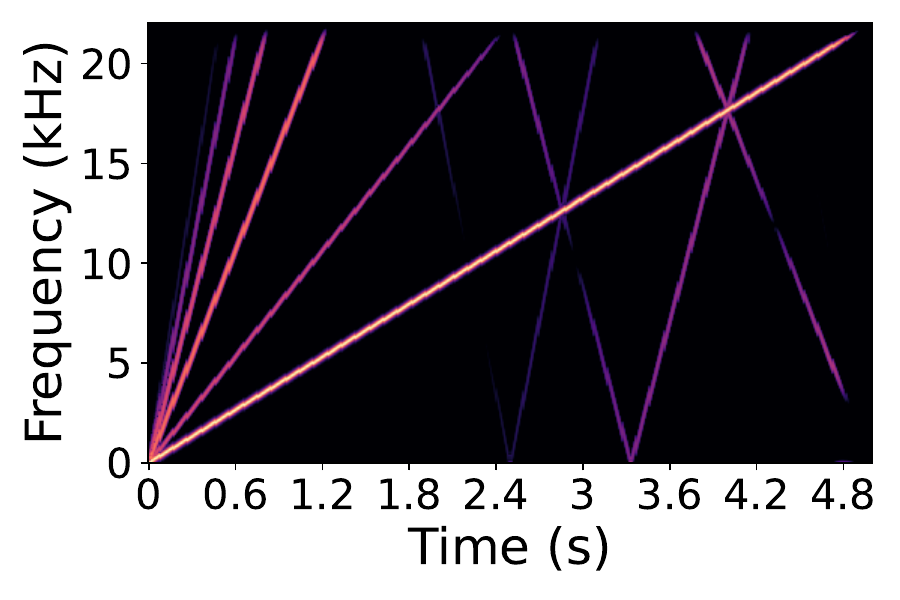}
         \caption{Ours}
         \label{fig:adaasnakebeta_2}
    \end{subfigure}
    \caption{Anti-aliasing case study by passing a sine sweep through different activations. ``O'' is the oversampling factor.}
    \vspace{-16pt}
    \label{fig:sine-sweep-case-study}
\end{figure*}

\begin{table}[t]
\begin{center}
  \centering
  \captionof{table}{AHR results of different activation functions and upsampling layers on the test signal benchmark.  ``O'' is the oversampling factor. The values are reported in dB scale. The best and the second best results of every column in each baseline setting are \textbf{bold} and \underline{underlined}.}
  \label{tab:results-test-signal-module}
  \setlength\tabcolsep{3pt}
  \resizebox{\linewidth}{!}{
  \begin{tabular}{lcccc}
    \toprule
    \multirow{2}{*}{\textbf{Module}} & \multicolumn{4}{c}{\textbf{AHR ($\downarrow$)}} \\ 
    \cmidrule(lr){2-5} & \textbf{Sine} & \textbf{Sawtooth} & \textbf{Triangle} & \textbf{Average} \\
    \midrule
    LeakyReLU & -17.68 & -36.70 & -21.39 & -25.25 \\
    ELU & -39.76 & -55.40 & \textbf{-39.20} & \underline{-44.79} \\
    SnakeBeta & -38.32 & -50.37 & -30.19 & -39.63 \\
    \revision{SnakeBeta ($O$ = 2)} & \revision{-38.67} & \revision{-54.92} & \revision{-30.36} & \revision{-41.32} \\
    \revision{SnakeBeta ($O$ = 4)} & \revision{-39.74} & \revision{-55.13} & \revision{-31.41} & \revision{-42.43} \\
    \revision{ADAA SnakeBeta} & \revision{\underline{-40.92}} & \revision{\underline{-57.84}} & \revision{-31.11} & \revision{-42.29} \\
    Ours & \textbf{-42.05} & \textbf{-58.33} & \underline{-37.47} & \textbf{-45.95} \\
    \midrule
    ConvTranspose & -24.28 & -16.11 & -24.62 & -21.67 \\
    Linear Interpolation & \textbf{-63.48} & \underline{-29.14} & \underline{-51.71} & \underline{-48.11} \\
    Nearest Interpolation & -34.20 & -14.83 & -28.54 & -25.86 \\
    Ours & \underline{-62.87} & \textbf{-39.92} & \textbf{-59.00} & \textbf{-53.93} \\
    \bottomrule
  \end{tabular}
  }
  \vspace{-20pt}
\end{center}
\end{table}

\subsubsection{Subjective Evaluation}

We conduct the MUSHRA~\cite{mushra} and Comparative Mean Opinion Score (C-MOS) tests for the subjective evaluation. For MUSHRA, 4 samples per domain are assessed. Listeners are asked to assign quality scores from 1 to 100 for each system. We use the ground-truth audio as the reference, and its 16\,kHz low-pass-filtered version as the hidden anchor. For C-MOS, 6 samples are evaluated. Listeners are asked to assign quality scores on a scale of -3 to 3 compared to the baseline. \revision{For the evaluation of singing voice, music, audio, and the industrial setting of speech, we invite 20 volunteers with experience in audio generation to participate in the test. For the academic speech setting, since quality perception is heavily influenced by semantics, we specifically recruited 5 native speakers per language via Prolific to assess their respective test sets and aggregated the final scores. To ensure data reliability, we integrated 2 attention-check trials in which listeners were instructed via audio to grade a sample within a specific range. Failure of either the attention-check or more than one hidden-anchor check resulted in exclusion.} All tests are conducted online, and listeners are instructed to use headphones in a quiet environment.

\subsection{Experimental Results}

\begin{figure}[t]
    \captionsetup[subfigure]{skip=3pt}
    \centering
    \begin{subfigure}[b]{0.48\columnwidth}
         \centering
         \includegraphics[width=\columnwidth, trim={0 0.4cm 0 0},clip]{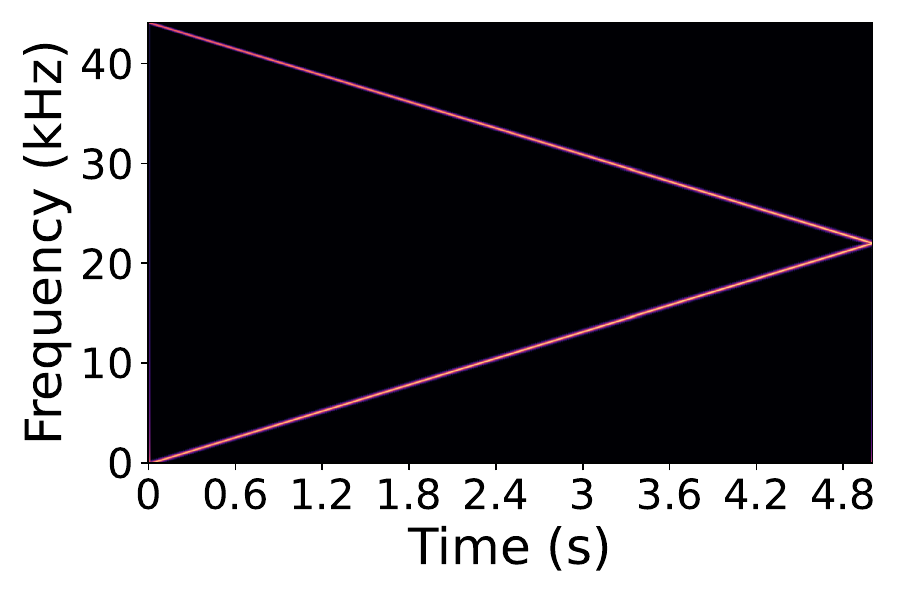}
         \caption{\revision{ConvTranspose}}
         \label{fig:convtranspose}
    \end{subfigure}
    \hfill
    \begin{subfigure}[b]{0.48\columnwidth}
         \centering
         \includegraphics[width=\columnwidth, trim={0 0.4cm 0 0},clip]{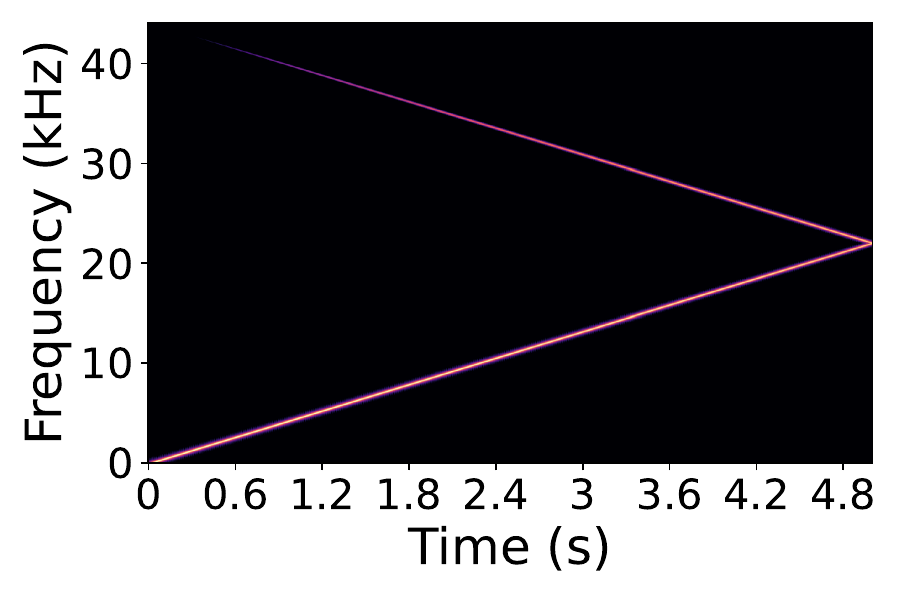}
         \caption{\revision{Linear Interpolation}}
         \label{fig:linear-interpolation-upsample}
    \end{subfigure}
    \\
    \begin{subfigure}[b]{0.48\columnwidth}
         \centering
         \includegraphics[width=\columnwidth, trim={0 0.4cm 0 0},clip]{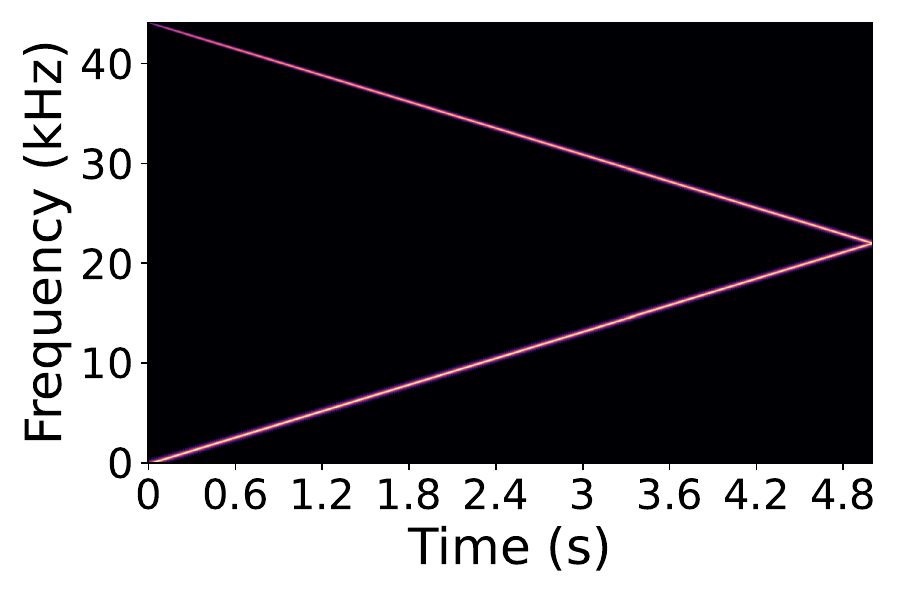}
         \caption{\revision{Nearest Interpolation}}
         \label{fig:nearest-interpolation-upsample}
    \end{subfigure}
    \hfill
    \begin{subfigure}[b]{0.48\columnwidth}
         \centering
         \includegraphics[width=\columnwidth, trim={0 0.4cm 0 0},clip]{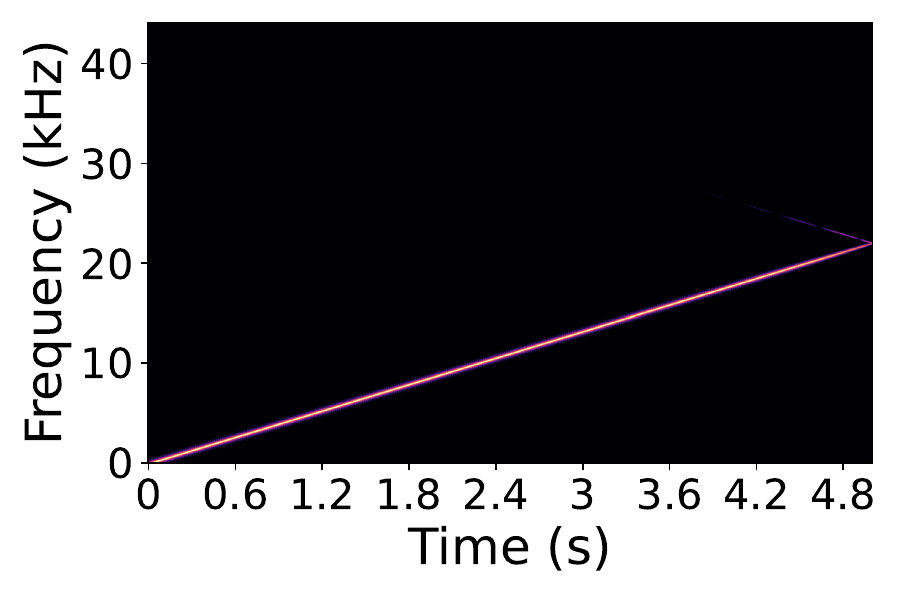}
         \caption{\revision{Ours}}
         \label{fig:ours}
    \end{subfigure}
    \caption{\revision{Anti-aliasing case study by passing a sine sweep through different upsampling layers.}}
    \vspace{-20pt}
    \label{fig:sine-sweep-case-study-upsampling}
\end{figure}

\subsubsection{Effectiveness of the Anti-Aliased Modules}

The objective evaluation results on the test signal benchmark are illustrated in Table~\ref{tab:results-test-signal-module}. For the activation function, LeakyReLU introduced the most aliasing artifacts, followed by SnakeBeta. \revision{Applying oversampling to SnakeBeta consistently mitigated aliasing artifacts, with higher oversampling factors yielding better results. Notably, our ADAA SnakeBeta without oversampling achieved anti-aliasing performance comparable to SnakeBeta with a $2\times$ oversampling factor. In contrast, ELU reduced aliasing artifacts by smoothing the sharp nonlinearity of ReLU with an exponential transition, performing slightly better than ADAA SnakeBeta. However, by combining ADAA SnakeBeta with a $2\times$ oversampling factor, our proposed method achieved the lowest average AHR, demonstrating its effectiveness.} For the upsampling layer, the randomly initialized ConvTranspose obtained the worst performance, as it introduced both the ``mirrored'' aliasing artifacts and ``tonal artifact''. Although the linear and nearest interpolations can mitigate these issues to some extent, their effectiveness remains limited, since they introduce additional ``filter artifact'' due to their poor filter frequency responses, as we discussed in Section~\ref{sec:analysis}.

To better illustrate the effectiveness of the proposed anti-aliased activation and upsampling modules, we conducted a case study by passing a sine sweep through different activation functions with various oversampling factors and upsampling layers, as shown in Figure~\ref{fig:sine-sweep-case-study} and Figure~\ref{fig:sine-sweep-case-study-upsampling}. \revision{For activation functions, it can be observed that both LeakyReLU and SnakeBeta introduces extensive amount of aliasing artifacts, followed by ELU, which generates noticeably fewer artifacts due to its smoother non-linearity. Meanwhile, oversampling proves to be a highly effective strategy; the aliasing artifacts produced by SnakeBeta visibly diminish as the oversampling factor increases from $2\times$ to $4\times$. Furthermore, ADAA is also beneficial for aliasing suppression, where ADAA SnakeBeta achieves similar performance compared to SnakeBeta with a $2\times$ oversampling factor, and ADAA SnakeBeta with a $2\times$ oversampling factor performs comparably to SnakeBeta with a $4\times$ oversampling factor, confirming its effectiveness.} \revision{For the upsampling layers, both ConvTranspose and nearest interpolation have almost no effect on aliasing suppression. While linear interpolation manages to remove a portion of the aliasing artifacts, a significant amount of residual artifacts still remains visible. Lastly, our proposed anti-aliased upsampling module successfully eliminates all the ``mirrored'' aliasing artifacts, demonstrating its superior effectiveness.}


\subsubsection{Effectiveness on Speech, Singing, Music, and Audio}

\begin{table*}[t]
\begin{center}
\caption{Analysis-synthesis results of different systems on speech and singing voice. The best and the second best results of every column (except those from Ground Truth) in each domain and baseline setting are \textbf{bold} and \underline{underlined}. ``Acad.'' means academic setting and ``Ind.'' means industrial setting. The MUSHRA scores are within 95\% Confidence Interval (CI).}
\label{tab:results-human-voice}
\setlength\tabcolsep{1.3pt}
\resizebox{\linewidth}{!}{
\begin{tabular}{clccccccccccc}

\toprule
\multirow{2}{*}{\textbf{Domain}} & \multirow{2}{*}{\textbf{System}} & \multicolumn{2}{c}{\revision{\textbf{RTF ($\downarrow$)}}} & \multirow{2}{*}{\textbf{\# Param}} & \multicolumn{2}{c}{\textbf{Mos-Pred ($\uparrow$)}} & \multicolumn{2}{c}{\textbf{F0RMSE ($\downarrow$)}} & \multicolumn{2}{c}{\textbf{Periodicity ($\downarrow$)}} & \multicolumn{2}{c}{\textbf{MUSHRA ($\uparrow$)}} \\ 
\cmidrule(lr){3-4} \cmidrule(lr){6-7} \cmidrule(lr){8-9} \cmidrule(lr){10-11} \cmidrule(lr){12-13}
& & \revision{\textbf{CPU}} & \revision{\textbf{GPU}} & & \textbf{Acad.} & \textbf{Ind.} & \textbf{Acad.} & \textbf{Ind.} & \textbf{Acad.} & \textbf{Ind.} & \textbf{Acad.} & \textbf{Ind.} \\
\midrule
\multirow{13}{*}{\textbf{Speech}}
& Ground Truth & \revision{/} & \revision{/} & / & 3.77 & 4.29 & 0.00 & 0.00 & 0.0000 & 0.0000 & \revision{80.56 $\pm$ 0.37} & \revision{89.73 $\pm$ 0.91} \\ 
\cmidrule(lr){2-13}
& Vocos & \revision{0.0163} & \revision{0.0005} & 14M & 3.22 & 4.08 & 68.67 & 40.09 & 0.0291 & 0.0084 & \revision{52.23 $\pm$ 0.51} & \revision{66.41 $\pm$ 2.45} \\ 
\cmidrule(lr){2-13}
& HiFi-GAN & \revision{0.4051} & \revision{0.0018} & 14M & 3.29 & 4.11 & 80.40 & 46.64 & 0.0296 & 0.0087 & \revision{50.91 $\pm$ 0.47} & \revision{67.36 $\pm$ 2.31} \\ 
& BigVGAN$_{\text{small}}$ & \revision{1.0278} & \revision{0.0078} & 14M & 3.39 & 4.08 & 61.35 & 40.96 & 0.0296 & 0.0115 & \revision{57.66 $\pm$ 0.49} & \revision{66.50 $\pm$ 2.19} \\ 
& BigVGAN$_{\text{large}}$ & \revision{4.2342} & \revision{0.0145} & 122M & 3.47 & 4.14 & \underline{50.30} & \underline{33.93} & \textbf{0.0246} & \textbf{0.0074} & \revision{63.43 $\pm$ 0.45} & \revision{68.23 $\pm$ 2.19} \\ 
& Pupu-Vocoder$_{\text{small}}$ & \revision{3.6171} & \revision{0.0124} & 14M & \underline{3.49} & \underline{4.23} & 58.25 & 37.28 & \underline{0.0272} & \underline{0.0080} & \revision{\underline{64.84 $\pm$ 0.45}} & \revision{\underline{74.27 $\pm$ 2.41}} \\ 
& Pupu-Vocoder$_{\text{large}}$ & \revision{18.8014} & \revision{0.0321} & 122M & \textbf{3.53} & \textbf{4.26} & \textbf{48.88} & \textbf{32.01} & 0.0290 & 0.0099 & \revision{\textbf{67.74 $\pm$ 0.46}} & \revision{\textbf{76.36 $\pm$ 2.16}} \\ 
\cmidrule(lr){2-13}
& EnCodec & \revision{0.5498} & \revision{0.0040} & 59M & 3.47 & 4.25 & 49.60 & 29.96 & 0.0206 & 0.0080 & \revision{64.28 $\pm$ 0.46} & \revision{69.00 $\pm$ 2.38} \\ 
& DAC & \revision{2.7722} & \revision{0.0040} & 154M & 3.60 & \textbf{4.29} & \textbf{32.93} & 24.46 & 0.0128 & 0.0044 & \revision{73.34 $\pm$ 0.42} & \revision{\textbf{88.14 $\pm$ 0.87}} \\ 
& BigCodec & \revision{6.8286} & \revision{0.0255} & 412M & \textbf{3.65} & \underline{4.28} & 35.27 & 23.73 & 0.0148 & 0.0043 & \revision{\textbf{76.62 $\pm$ 0.40}} & \revision{86.45 $\pm$ 0.90} \\ 
& Pupu-Codec$_{\text{small}}$ & \revision{2.2569} & \revision{0.0069} & 32M & 3.54 & 4.25 & 38.83 & \underline{23.47} & \underline{0.0116} & \textbf{0.0037} & \revision{74.15 $\pm$ 0.42} & \revision{77.68 $\pm$ 2.12} \\ 
& Pupu-Codec$_{\text{large}}$ & \revision{10.2016} & \revision{0.0157} & 119M & \underline{3.61} & \textbf{4.29} & \underline{35.15} & \textbf{22.32} & \textbf{0.0111} & \underline{0.0040} & \revision{\underline{74.97 $\pm$ 0.42}} & \revision{\underline{86.51 $\pm$ 0.99}} \\ 
\midrule
\multirow{13}{*}{\makecell{\textbf{Singing}\\ \textbf{Voice}}}
& Ground Truth & \revision{/} & \revision{/} & / & 4.18 & 4.33 & 0.00 & 0.00 & 0.0000 & 0.0000 & 86.13 $\pm$ 0.82 & 89.30 $\pm$ 0.74 \\ 
\cmidrule(lr){2-13}
& Vocos & \revision{0.0163} & \revision{0.0005} & 14M & 3.66 & 3.61 & 32.56 & 25.22 & 0.0933 & 0.1112 & 49.84 $\pm$ 1.38 & 42.92 $\pm$ 1.23 \\ 
\cmidrule(lr){2-13}
& HiFi-GAN & \revision{0.4051} & \revision{0.0019} & 14M & 3.84 & 3.93 & 32.67 & 22.59 & 0.0962 & 0.1211 & 56.29 $\pm$ 1.17 & 57.68 $\pm$ 1.09 \\ 
& BigVGAN$_{\text{small}}$ & \revision{1.0278} & \revision{0.0078} & 14M & 3.76 & 3.78 & 32.07 & 21.34 & 0.0949 & 0.1159 & 51.26 $\pm$ 1.24 & 49.19 $\pm$ 1.19 \\ 
& BigVGAN$_{\text{large}}$ & \revision{4.2342} & \revision{0.0144} & 122M & 3.88 & 3.92 & \underline{26.56} & \underline{17.92} & \textbf{0.0754} & \textbf{0.0912} & 58.28 $\pm$ 1.26 & 53.08 $\pm$ 1.14 \\ 
& Pupu-Vocoder$_{\text{small}}$ & \revision{3.6171} & \revision{0.0124} & 14M & \underline{4.08} & \underline{4.22} & 29.17 & 19.13 & 0.0839 & 0.1110 & \underline{68.26 $\pm$ 1.20} & \underline{65.73 $\pm$ 1.25} \\ 
& Pupu-Vocoder$_{\text{large}}$ & \revision{18.8014} & \revision{0.0321} & 122M & \textbf{4.09} & \textbf{4.24} & \textbf{25.86} & \textbf{17.12} & \underline{0.0827} & \underline{0.1011} & \textbf{70.05 $\pm$ 1.14} & \textbf{70.84 $\pm$ 1.15} \\ 
\cmidrule(lr){2-13}
& EnCodec & \revision{0.5498} & \revision{0.0040} & 59M & 3.97 & 4.19 & 26.72 & 15.76 & 0.0875 & 0.1349 & 54.74 $\pm$ 1.41 & 57.16 $\pm$ 1.26 \\ 
& DAC & \revision{2.7722} & \revision{0.0040} & 154M & \underline{4.16} & \textbf{4.33} & \textbf{19.57} & 12.38 & \underline{0.0557} & 0.0904 & 80.55 $\pm$ 0.99 & \underline{85.43 $\pm$ 0.85} \\ 
& BigCodec & \revision{6.8286} & \revision{0.0255} & 412M & \textbf{4.17} & \underline{4.32} & 20.44 & \underline{12.37} & 0.0598 & \underline{0.0860} & \underline{82.74 $\pm$ 0.84} & 84.81 $\pm$ 0.83 \\ 
& Pupu-Codec$_{\text{small}}$ & \revision{2.2569} & \revision{0.0069} & 32M & 4.10 & 4.31 & 22.31 & 13.70 & 0.0559 & 0.0979 & 75.61 $\pm$ 1.28 & 78.97 $\pm$ 1.16 \\ 
& Pupu-Codec$_{\text{large}}$ & \revision{10.2016} & \revision{0.0157} & 119M & \textbf{4.17} & \textbf{4.33} & \underline{19.84} & \textbf{12.34} & \textbf{0.0528} & \textbf{0.0834} & \textbf{83.79 $\pm$ 0.94} & \textbf{85.65 $\pm$ 0.86} \\ 
\bottomrule

\end{tabular}
}
\vspace{-6pt}
\end{center}
\end{table*}

\begin{table*}[htbp]
\begin{center}
\caption{Analysis-synthesis results of different systems on music and audio. The best and the second best results of every column (except those from Ground Truth) in each domain and baseline setting are \textbf{bold} and \underline{underlined}. ``Acad.'' means academic setting and ``Ind.'' means industrial setting. The MUSHRA scores are within 95\% Confidence Interval (CI).}
\label{tab:results-audio}
\setlength\tabcolsep{1.3pt}
\resizebox{\linewidth}{!}{
\begin{tabular}{clccccccccccc}

\toprule
\multirow{2}{*}{\textbf{Domain}} & \multirow{2}{*}{\textbf{System}} & \multicolumn{2}{c}{\revision{\textbf{RTF ($\downarrow$)}}} & \multirow{2}{*}{\textbf{\# Param}} & \multicolumn{2}{c}{\textbf{FAD ($\downarrow$)}} & \multicolumn{2}{c}{\textbf{M-STFT ($\downarrow$)}} & \multicolumn{2}{c}{\textbf{ViSQOL ($\uparrow$)}} & \multicolumn{2}{c}{\textbf{MUSHRA ($\uparrow$)}} \\ 
\cmidrule(lr){3-4} \cmidrule(lr){6-7} \cmidrule(lr){8-9} \cmidrule(lr){10-11} \cmidrule(lr){12-13}
& & \revision{\textbf{CPU}} & \revision{\textbf{GPU}} & & \textbf{Acad.} & \textbf{Ind.} & \textbf{Acad.} & \textbf{Ind.} & \textbf{Acad.} & \textbf{Ind.} & \textbf{Acad.} & \textbf{Ind.} \\
\midrule
\multirow{13}{*}{\makecell{\textbf{Music}}}
& Ground Truth & \revision{/} & \revision{/} & / & 0.000 & 0.000 & 0.00 & 0.00 & 5.00 & 5.00 & 87.24 $\pm$ 0.85 & 86.48 $\pm$ 1.26 \\ 
\cmidrule(lr){2-13}
& Vocos & \revision{0.0163} & \revision{0.0005} & 14M & 0.017 & 0.037 & 0.72 & 0.83 & 4.54 & 4.42 & 58.89 $\pm$ 1.42 & 48.00 $\pm$ 1.69 \\
\cmidrule(lr){2-13}
& HiFi-GAN & \revision{0.4051} & \revision{0.0018} & 14M & 0.044 & 0.085 & 0.83 & 0.90 & 4.33 & 4.32 & 54.84 $\pm$ 1.28 & 37.74 $\pm$ 1.57 \\ 
& BigVGAN$_{\text{small}}$ & \revision{1.0278} & \revision{0.0078} & 14M & 0.021 & 0.054 & 0.79 & \underline{0.88} & 4.55 & 4.46 & 58.43 $\pm$ 1.17 & 42.55 $\pm$ 1.60 \\ 
& BigVGAN$_{\text{large}}$ & \revision{4.2342} & \revision{0.0145} & 122M & \textbf{0.014} & \textbf{0.044} & 0.82 & 0.93 & \textbf{4.61} & \textbf{4.54} & \underline{65.68 $\pm$ 1.34} & \underline{50.42 $\pm$ 1.90} \\  
& Pupu-Vocoder$_{\text{small}}$ & \revision{3.6171} & \revision{0.0124} & 14M & 0.043 & 0.087 & \underline{0.75} & 0.91 & 4.37 & 4.35 & 56.35 $\pm$ 1.43 & 42.87 $\pm$ 1.90 \\ 
& Pupu-Vocoder$_{\text{large}}$ & \revision{18.8014} & \revision{0.0321} & 122M & \underline{0.017} & \underline{0.049} & \textbf{0.71} & \textbf{0.83} & \underline{4.60} & \underline{4.48} & \textbf{70.38 $\pm$ 1.26} & \textbf{56.42 $\pm$ 1.66} \\ 
\cmidrule(lr){2-13}
& EnCodec & \revision{0.5498} & \revision{0.0040} & 59M & 0.141 & 0.136 & 0.88 & 0.95 & 4.05 & 4.23 & 52.65 $\pm$ 2.53 & 39.97 $\pm$ 1.79 \\ 
& DAC & \revision{2.7722} & \revision{0.0040} & 154M & 0.040 & 0.045 & \underline{0.76} & \textbf{0.83} & 4.31 & \underline{4.43} & 71.78 $\pm$ 1.24 & 72.65 $\pm$ 1.50 \\ 
& BigCodec & \revision{6.8286} & \revision{0.0255} & 412M & \underline{0.033} & \textbf{0.029} & 0.86 & \underline{0.88} & \underline{4.32} & \textbf{4.44} & \underline{72.87 $\pm$ 1.18} & \underline{73.09 $\pm$ 1.25} \\ 
& Pupu-Codec$_{\text{small}}$ & \revision{2.2569} & \revision{0.0069} & 32M & 0.036 & 0.066 & 0.78 & 0.90 & 4.12 & 4.29 & 68.16 $\pm$ 1.40 & 65.23 $\pm$ 1.60 \\ 
& Pupu-Codec$_{\text{large}}$ & \revision{10.2016} & \revision{0.0157} & 119M & \textbf{0.019} & \underline{0.033} & \textbf{0.75} & \textbf{0.83} & \textbf{4.34} & \textbf{4.44} & \textbf{73.14 $\pm$ 1.27} & \textbf{74.39 $\pm$ 1.40} \\ 
\midrule
\multirow{13}{*}{\textbf{Audio}}
& Ground Truth & \revision{/} & \revision{/} & / & 0.000 & 0.000 & 0.00 & 0.00 & 5.00 & 5.00 & 88.22 $\pm$ 0.95 & 82.83 $\pm$ 1.02 \\ 
\cmidrule(lr){2-13}
& Vocos & \revision{0.0163} & \revision{0.0005} & 14M & 0.022 & 0.018 & 0.88 & 0.84 & 4.50 & 4.55 & 74.50 $\pm$ 1.30 & 64.25 $\pm$ 1.37 \\ 
\cmidrule(lr){2-13}
& HiFi-GAN & \revision{0.4051} & \revision{0.0019} & 14M & 0.048 & 0.037 & 0.88 & 0.95 & 4.23 & 4.38 & 63.59 $\pm$ 1.53 & 58.78 $\pm$ 1.52 \\ 
& BigVGAN$_{\text{small}}$ & \revision{1.0278} & \revision{0.0078} & 14M & 0.019 & 0.020 & \underline{0.84} & 0.91 & 4.46 & 4.56 & 71.97 $\pm$ 1.45 & 66.11 $\pm$ 1.50 \\ 
& BigVGAN$_{\text{large}}$ & \revision{4.2342} & \revision{0.0144} & 122M & \textbf{0.013} & \textbf{0.017} & 0.97 & 0.95 & \textbf{4.53} & \textbf{4.61} & \underline{76.66 $\pm$ 1.35} & \underline{73.17 $\pm$ 1.41} \\ 
& Pupu-Vocoder$_{\text{small}}$ & \revision{3.6171} & \revision{0.0124} & 14M & 0.031 & 0.032 & 0.88 & \underline{0.90} & 4.24 & 4.39 & 71.47 $\pm$ 1.43 & 65.28 $\pm$ 1.50 \\ 
& Pupu-Vocoder$_{\text{large}}$ & \revision{18.8014} & \revision{0.0321} & 122M & \underline{0.017} & \underline{0.018} & \textbf{0.83} & \textbf{0.86} & \underline{4.47} & \underline{4.58} & \textbf{77.84 $\pm$ 1.22} & \textbf{73.47 $\pm$ 1.23} \\ 
\cmidrule(lr){2-13}
& EnCodec & \revision{0.5498} & \revision{0.0040} & 59M & 0.207 & 0.089 & 1.08 & 1.07 & 3.91 & 3.96 & 55.56 $\pm$ 1.54 & 45.36 $\pm$ 1.50 \\  
& DAC & \revision{2.7722} & \revision{0.0040} & 154M & 0.087 & 0.049 & 0.91 & \underline{0.92} & 4.14 & 4.25 & 74.56 $\pm$ 1.51 & 70.56 $\pm$ 1.18 \\ 
& BigCodec & \revision{6.8286} & \revision{0.0255} & 412M & 0.079 & \underline{0.042} & 1.03 & 0.99 & \underline{4.18} & \underline{4.27} & \underline{75.00 $\pm$ 1.28} & \underline{73.75 $\pm$ 1.08} \\  
& Pupu-Codec$_{\text{small}}$ & \revision{2.2569} & \revision{0.0069} & 32M & \underline{0.072} & 0.060 & \underline{0.90} & 0.93 & 3.98 & 4.02 & 69.28 $\pm$ 1.56 & 67.86 $\pm$ 1.37 \\ 
& Pupu-Codec$_{\text{large}}$ & \revision{10.2016} & \revision{0.0157} & 119M & \textbf{0.046} & \textbf{0.039} & \textbf{0.88} & \textbf{0.90} & \textbf{4.19} & \textbf{4.29} & \textbf{75.59 $\pm$ 1.36} & \textbf{75.00 $\pm$ 1.09} \\ 
\bottomrule

\end{tabular}
}
\end{center}
\end{table*}

We run experiments on speech, singing voice, music, and audio to show the effectiveness of our proposed models. The evaluation results are illustrated in Table~\ref{tab:results-human-voice} and Table~\ref{tab:results-audio}. 
Generally, TFR-based models perform indeed worse than time-domain models, as explicitly modeling the phase is inherently difficult.
For speech, regarding neural vocoders, both Pupu-Vocoder models have better scores on MOS-Pred and F0RMSE. 
While they have slightly higher values on Periodicity, their MUSHRA scores consistently outperform all baselines, confirming their effectiveness; for neural codecs, Pupu-Codec$_{\text{small}}$ outperforms Encodec and achieves performance on par with significantly larger models, showing its parameter efficiency. 
Meanwhile, Pupu-Codec$_{\text{large}}$ shows comparable performance to baselines across both objective and subjective metrics, validating its effectiveness.
For singing voice, a similar conclusion can be drawn from Pupu-Vocoder models and Pupu-Codec$_{\text{small}}$ as in speech. For Pupu-Codec$_{\text{large}}$, although it yields comparable objective results to the baselines, subjective evaluation reveals its superior performance, confirming the effectiveness of the anti-aliased modules.
For music and audio, regarding neural vocoders, Pupu-Vocoder$_{\text{small}}$ performs on par with HiFi-GAN and BigVGAN$_{\text{small}}$, and Pupu-Vocoder$_{\text{large}}$ yields comparable FAD and VisQOL with better M-STFT scores, while having significantly better performance on MUSHRA, illustrating its effectiveness; regarding neural codecs, the Pupu-Codec$_{\text{small}}$ model outperforms Encodec and achieves performance on par with other baselines, while Pupu-Codec$_{\text{large}}$ consistently outperform all the baseline systems both objectively and subjectively, showing its superior synthesis quality. \rerevision{Lastly, the RTF scores indicate that while our proposed models exhibit higher CPU computational cost due to the absence of kernel optimizations for DDSP operations, specifically oversampling and filtering, their GPU computational cost remains in an acceptable range. For instance, Pupu-Codec$_{\text{large}}$ is slower than BigCodec on the CPU (10.2016 vs 6.8286); however, because these DDSP operations are well-optimized on the GPU, Pupu-Codec$_{\text{large}}$ runs faster there (0.0157 vs 0.0255), given its fewer parameters (119M vs 412M).} As a result, since our methods are targeted at offline production scenarios (e.g., Vocaloid and Voiceroid), where producers accept a higher computational time in exchange for superior synthesis quality, the overall computational cost for our proposed systems is acceptable. \rerevision{Note that we deliberately evaluate both \textit{small} and \textit{large} settings of our proposed models to accommodate diverse practical needs regarding different computational cost tolerances.}

We further conducted a case study on singing voice to illustrate the effectiveness of our proposed models, as shown in the Figure~\ref{fig:singing-voice-case-study}. As we can see, the Vocos and BigGAN models are unable to generate reasonable harmonic structures at such a high frequency, instead producing noise. \revision{Specifically, regarding Vocos, it operates on the TF domain, and accurately recovering the phase information in the high-frequency band is exceedingly difficult. Thus, the degraded phase hinders the reconstruction of fine-grained high-frequency details, resulting in metallic-sounding ``straight lines''. Regarding BigVGAN, its unconstrained nonlinear activation function and upsampling layers introduce excessive aliasing artifacts. In the high-frequency bands, the introduced aliasing artifacts cause phase cancellation, thereby collapsing the harmonic structure into unstructured noise.} In contrast, the DAC and BigCodec models can generate harmonic components in high-frequency bands but also introduce significant aliased components, resulting in a blurred, noisy high-frequency region that degrades synthesis quality, similar to BigVGAN. Unlike these baseline models, our proposed models can visibly generate harmonics in high-frequency bands without introducing aliasing artifacts. In particular, comparing our internal model variants, the Pupu-Vocoder models tend to generate harmonic components with distorted shapes, with the smaller variant exhibiting noticeably more distortions than the larger one, which may be due to their lack of implicit phase modeling. Furthermore, the Pupu-Codec$_\text{small}$ model can generate harmonic components with the correct shape but tends to overproduce harmonics that do not appear in the original waveform, resulting in hissing noises in the background, which may be because it lacks sufficient parameters in the final layers to effectively filter out these unwanted harmonics. In contrast, the Pupu-Codec$_\text{large}$ model can accurately reconstruct the full harmonic structure, benefiting from its increased model capacity. In summary, this visual case study confirms the effectiveness of our proposed models in achieving high-fidelity neural audio synthesis.

\begin{table*}[t]
\begin{center}
\caption{Analysis-synthesis results of different systems on singing voice in different bitrates. The best and the second best results of every column in each domain and bitrate setting are \textbf{bold} and \underline{underlined}. ``Acad.'' means academic setting and ``Ind.'' means industrial setting. The MUSHRA scores are within 95\% Confidence Interval (CI).}
\label{tab:results-singing-voice-bitrate}
\setlength\tabcolsep{4pt}
\resizebox{\linewidth}{!}{
\begin{tabular}{clcccccccccc}

\toprule
\multirow{2}{*}{\textbf{Bitrate}} & \multirow{2}{*}{\textbf{System}} & 
\multicolumn{2}{c}{\revision{\textbf{RTF ($\downarrow$)}}} & \multirow{2}{*}{\textbf{\# Param}} & \multicolumn{2}{c}{\textbf{Mos-Pred ($\uparrow$)}} & \multicolumn{2}{c}{\textbf{F0RMSE ($\downarrow$)}} & \multicolumn{2}{c}{\textbf{Periodicity ($\downarrow$)}} & \multirow{2}{*}{\textbf{MUSHRA ($\uparrow$)}} \\ 
\cmidrule(lr){3-4} \cmidrule(lr){6-7} \cmidrule(lr){8-9} \cmidrule(lr){10-11} 
& & \revision{\textbf{CPU}} & \revision{\textbf{GPU}} & & \textbf{Acad.} & \textbf{Ind.} & \textbf{Acad.} & \textbf{Ind.} & \textbf{Acad.} & \textbf{Ind.} & \\
\midrule
/ & Ground Truth & \revision{/} & \revision{/} & / & 4.18 & 4.33 & 0.00 & 0.00 & 0.0000 & 0.0000 & 86.78 $\pm$ 0.23 \\ 
\midrule
\multirow{5}{*}{\textbf{8\,kbps}}
& EnCodec & \revision{0.5498} & \revision{0.0040} & 59M & 3.97 & 4.19 & 26.72 & 15.76 & 0.0875 & 0.1349 & 56.81 $\pm$ 1.59 \\ 
& DAC & \revision{2.7722} & \revision{0.0040} & 154M & \underline{4.16} & \textbf{4.33} & \textbf{19.57} & 12.38 & \underline{0.0557} & 0.0904 & \underline{81.39 $\pm$ 1.07} \\ 
& BigCodec & \revision{6.8286} & \revision{0.0255} & 412M & \textbf{4.17} & \underline{4.32} & 20.44 & \underline{12.37} & 0.0598 & \underline{0.0860} & 81.14 $\pm$ 1.16 \\ 
& Pupu-Codec$_{\text{small}}$ & \revision{2.2569} & \revision{0.0069} & 32M & 4.10 & 4.31 & 22.31 & 13.70 & 0.0559 & 0.0979 & 72.97 $\pm$ 1.38 \\ 
& Pupu-Codec$_{\text{large}}$ & \revision{10.2016} & \revision{0.0157} & 119M & \textbf{4.17} & \textbf{4.33} & \underline{19.84} & \textbf{12.34} & \textbf{0.0528} & \textbf{0.0834} & \textbf{82.64 $\pm$ 1.09} \\ 
\midrule
\multirow{5}{*}{\textbf{5.33\,kbps}}
& EnCodec & \revision{0.5498} & \revision{0.0040} & 59M & 3.82 & 4.08 & 30.79 & 18.04 & 0.0993 & 0.1417 & 52.61 $\pm$ 1.44 \\
& DAC & \revision{2.7722} & \revision{0.0040} & 154M & \textbf{4.15} & \textbf{4.32} & \textbf{21.50} & \textbf{13.84} & \underline{0.0630} & \underline{0.0974} & \underline{79.22 $\pm$ 1.09} \\ 
& BigCodec & \revision{6.8286} & \revision{0.0255} & 412M & \underline{4.14} & \underline{4.30} & 24.05 & 14.65 & 0.0719 & 0.1008 & 74.25 $\pm$ 1.13  \\ 
& Pupu-Codec$_{\text{small}}$ & \revision{2.2569} & \revision{0.0069} & 32M & 4.05 & 4.28 & 25.52 & 16.64 & 0.0667 & 0.1063 & 73.92 $\pm$ 1.37 \\ 
& Pupu-Codec$_{\text{large}}$ & \revision{10.2016} & \revision{0.0157} & 119M & \textbf{4.15} & \textbf{4.32} & \underline{23.53} & \underline{14.38} & \textbf{0.0624} & \textbf{0.0941} & \textbf{81.56 $\pm$ 0.87} \\
\midrule
\multirow{5}{*}{\textbf{2.67\,kbps}}
& EnCodec & \revision{0.5498} & \revision{0.0040} & 59M & 3.34 & 3.64 & 44.99 & 26.10 & 0.1410 & 0.1673 & 42.57 $\pm$ 1.36 \\ 
& DAC & \revision{2.7722} & \revision{0.0040} & 154M & \textbf{4.06} & \textbf{4.27} & \textbf{27.05} & \textbf{18.24} & \textbf{0.0825} & \textbf{0.1182} & \underline{72.68 $\pm$ 0.97} \\ 
& BigCodec & \revision{6.8286} & \revision{0.0255} & 412M & \underline{4.02} & 4.23 & 30.37 & 20.19 & 0.0930 & 0.1250 & 70.43 $\pm$ 1.29 \\ 
& Pupu-Codec$_{\text{small}}$ & \revision{2.2569} & \revision{0.0069} & 32M & 3.87 & 4.17 & 32.64 & 21.87 & 0.0901 & 0.1279 & 67.51 $\pm$ 1.39 \\ 
& Pupu-Codec$_{\text{large}}$ & \revision{10.2016} & \revision{0.0157} & 119M & \textbf{4.06} & \underline{4.25} & \underline{29.84} & \underline{19.48} & \underline{0.0841} & \underline{0.1195} & \textbf{73.97 $\pm$ 1.22} \\
\midrule
\multirow{5}{*}{\textbf{1.78\,kbps}}
& EnCodec & \revision{0.5498} & \revision{0.0040} & 59M & 2.64 & 2.84 & 64.44 & 51.13 & 0.1967 & 0.2277 & 25.94 $\pm$ 1.22 \\ 
& DAC & \revision{2.7722} & \revision{0.0040} & 154M & \textbf{3.94} & \textbf{4.18} & \textbf{33.34} & \textbf{22.53} & \textbf{0.1005} & \textbf{0.1343} & \textbf{67.53 $\pm$ 1.21} \\ 
& BigCodec & \revision{6.8286} & \revision{0.0255} & 412M & 3.84 & 4.08 & 36.35 & 25.76 & 0.1051 & 0.1437 & 55.00 $\pm$ 1.28 \\ 
& Pupu-Codec$_{\text{small}}$ & \revision{2.2569} & \revision{0.0069} & 32M & 3.65 & 3.95 & 39.16 & 28.31 & 0.1133 & 0.1521 & 53.33 $\pm$ 1.29 \\ 
& Pupu-Codec$_{\text{large}}$ & \revision{10.2016} & \revision{0.0157} & 119M & \underline{3.87} & \underline{4.11} & \underline{35.51} & \underline{23.61} & \underline{0.1018} & \underline{0.1365} & \underline{65.17 $\pm$ 1.18} \\

\bottomrule

\end{tabular}
}
\end{center}
\vspace{-12pt}
\end{table*}

\subsubsection{Effectiveness on Dynamic Bitrate Encoding}

We also explored the performance of our proposed Pupu-Codec models under the dynamic bitrate encoding scenario, \revision{as illustrated in Table~\ref{tab:results-singing-voice-bitrate}}.
\revision{To achieve dynamic bitrate encoding, we adjust the number of active quantization layers within the RVQ module while keeping other parameters, such as the encoding frame rate and codebook size, strictly unmodified. We apply a random-codebook dropout strategy during training to ensure that all models can reconstruct the audio using an arbitrary number of layers.} 
We perform our evaluation in the industrial singing voice setting, as its rich harmonic structure and higher modeling difficulty make it easier to distinguish the quality of different models.
Specifically, while the Encodec model exhibits severe quality degradation in low-bitrate conditions, our proposed PupuCodec$_\text{small}$ model maintains performance and is comparable to the DAC and BigCodec models, despite having fewer parameters.
Meanwhile, our Pupu-Codec$_\text{large}$ model exhibits comparable objective metrics to those of DAC and BigCodec models. 
Moreover, in subjective evaluation, it outperforms the DAC and BigCodec models in high and medium-bitrate scenarios (8\,kbps, 5.33\,kbps, and 2.67\,kbps) and is on par with them in low-bitrate scenarios (1.78\,kbps).
We hypothesize that this convergence occurs because, at such restricted bitrates, the information bottleneck limits the reconstruction of high-frequency components, for which our approach is particularly effective.
Nevertheless, despite this challenge, our proposed Pupu-Codec models still achieve remarkable performance compared to the baseline models.

\subsubsection{Ablation Study}

We conducted an ablation study on singing voice to illustrate the effectiveness of our proposed anti-aliased modules, as shown in Table~\ref{tab:results-ablation}. 
\revision{Specifically, we isolate the impact of each component by keeping all other model components strictly unmodified while either removing the target module or replacing it with an alternative one}.
For activation functions, oversampling is crucial for audio quality; although it introduces extra computational cost, removing it degrades synthesis quality both objectively and subjectively. 
Meanwhile, their anti-aliasing abilities also correlate with the synthesis quality.
Among them, our ADAA SnakeBeta achieves the best result both objectively and subjectively, followed by the SnakeBeta and ELU. 
For different upsampling layers, the ConvTranspose, linear, and nearest interpolation achieve similar results objectively. 
The linear and nearest interpolation methods achieve the best subjective performance, as they do not exhibit ``tonal artifacts'' and attenuate part of the ``mirrored'' aliasing artifacts compared to the ConvTranspose layer. 
Finally, the deterministic noise prior proves to be a crucial factor in achieving synthesis fidelity; removing it would greatly degrade the synthesis performance both objectively and subjectively. 
This aligns with previous work~\cite{noise-prior}, which demonstrates that standard neural networks exhibit spectral bias, hindering their ability to reconstruct high-frequency details. This issue can be resolved by projecting inputs into a high-frequency prior, thus overcoming the low-frequency bias. Overall, the experimental results validate our proposed anti-aliased modules, illustrating their effectiveness.


\begin{table*}[t]
\begin{center}
\caption{Ablation experiment results on singing voice. The best and the second best results of every column in each ablation setting are \textbf{bold} and \underline{underlined}. ``Acad.'' means academic setting, ``Ind.'' means industrial setting, ``w/o'' means ``without'', and ``$\flatMapsto$'' means ``replace''. The C-MOS scores are within 95\% Confidence Interval (CI).}
\label{tab:results-ablation}
\resizebox{\linewidth}{!}{
\begin{tabular}{lccccccccc}

\toprule
\multirow{2}{*}{\textbf{System}} & \multicolumn{2}{c}{\revision{\textbf{RTF ($\downarrow$)}}} & \multicolumn{2}{c}{\textbf{Mos-Pred ($\uparrow$)}} & \multicolumn{2}{c}{\textbf{F0RMSE ($\downarrow$)}} & \multicolumn{2}{c}{\textbf{Periodicity ($\downarrow$)}} & \multirow{2}{*}{\textbf{C-MOS ($\uparrow$)}} \\ 
\cmidrule(lr){2-3} \cmidrule(lr){4-5} \cmidrule(lr){6-7} \cmidrule(lr){8-9} 
& \revision{\textbf{CPU}} & \revision{\textbf{GPU}} & \textbf{Acad.} & \textbf{Ind.} & \textbf{Acad.} & \textbf{Ind.} & \textbf{Acad.} & \textbf{Ind.} & \\
\midrule
Pupu-Vocoder$_{\text{small}}$ & \revision{3.6171} & \revision{0.0124} & \textbf{4.081} & \textbf{4.221} & \textbf{29.17} & \textbf{19.13} & \textbf{0.0839} & \textbf{0.1110} & / \\ 
\midrule
w/o Oversampling & \revision{1.9003} & \revision{0.0060} & 3.991 & 4.133 & 33.40 & 27.04 & 0.0988 & 0.1351 & \underline{-0.38 $\pm$ 0.02} \\ 
Ours $\flatMapsto$ LeakyReLU & \revision{1.4252} & \revision{0.0074} & 3.839 & 3.940 & 34.83 & 26.82 & 0.1060 & 0.1452 & -1.45 $\pm$ 0.02 \\ 
Ours $\flatMapsto$ ELU & \revision{1.6354} & \revision{0.0074} & 3.972 & 4.075 & 34.85 & 23.85 & 0.1019 & \underline{0.1297} & -0.93 $\pm$ 0.02 \\ 
Ours $\flatMapsto$ SnakeBeta & \revision{1.9338} & \revision{0.0089} & \underline{4.047} & \underline{4.153} & \underline{30.97} & \underline{20.11} & \underline{0.0934} & 0.1360 & -0.52 $\pm$ 0.02 \\ 
\midrule
w/o Deterministic Prior & \revision{3.1338} & \revision{0.0119} & 3.811 & 3.896 & 35.35 & 28.14 & 0.1059 & 0.1488 & -1.49 $\pm$ 0.02 \\ 
Ours $\flatMapsto$ ConvTranspose & \revision{3.0314} & \revision{0.0111} & 4.040 & 4.182 & 29.35 & 20.76 & 0.0871 & \underline{0.1141} & -0.15 $\pm$ 0.02 \\ 
Ours $\flatMapsto$ Linear Interpolation & \revision{2.8818} & \revision{0.0112} & 4.042 & \underline{4.186} & \underline{29.21} & \underline{19.67} & 0.0883 & 0.1178 & \underline{-0.01 $\pm$ 0.02} \\ 
Ours $\flatMapsto$ Nearest Interpolation & \revision{2.8413} & \revision{0.0110} & \underline{4.044} & 4.170 & 29.61 & 19.82 & \underline{0.0843} & 0.1148 & -0.04 $\pm$ 0.02 \\ 
\bottomrule

\end{tabular}
}
\vspace{-12pt}
\end{center}
\end{table*}

\section{Conclusion}

This paper addresses the synthesis fidelity limitations in upsampling-based neural audio generation that are introduced by inadequately designed model architectures. By analyzing and identifying the sources of ``folded-back'' and ``mirrored'' aliasing artifacts, as well as the ``tonal artifact'', we propose Pupu-Vocoder and Pupu-Codec, which incorporate our novel anti-aliased activation and upsampling modules. Experimental results demonstrate that our proposed models can consistently outperform existing baselines across singing voice, music, and audio, while yielding comparable results on speech.

\section{Future Works}

Although our proposed Pupu-Vocoder and Pupu-Codec demonstrate superior performance across multiple domains, several issues remain to be addressed for future research. Firstly, the Pupu-Codec's performance in extremely low-bitrate scenarios needs to be optimized, as reducing the number of audio tokens remains essential in large language model (LLM)-based acoustic models. Second, the overall synthesis quality in music generation remains limited. Thus, exploring advanced architectural designs to better capture polyphonic harmonic structures is needed. Finally, a notable limitation of our current models is the high computational cost introduced by the oversampling strategy. To address this problem, we plan to investigate the adaptation of higher-order ADAA techniques to achieve better anti-aliasing performance, thereby eliminating the need for additional oversampling.

\section{Acknowledgment}

This work is a joint effort by Spellbrush, Aalto University, and The Chinese University of Hong Kong, Shenzhen. We extend our sincere gratitude to everyone who participated in discussions and provided valuable insights throughout the development of this work. Finally, the first author would like to acknowledge her partner, Zeyu Dou, for his consistent support throughout her life and career. In recognition of his fondness for rabbits, \textit{Pupu}, known as ``bunny'' in Finnish, was adopted as the prefix for the proposed models.

\section{Generative AI Use Disclosure}
In accordance with the IEEE SPS policy on the use of LLM, the authors disclose the use of Gemini 3 Pro in the preparation of this manuscript. The tool was strictly used to improve language and clarity and to accelerate code development. The authors confirm that no significant components or entire sections of this manuscript were generated by the LLM. All AI-assisted outputs, including text and code, have been thoroughly reviewed, edited, and verified for accuracy. The authors take full responsibility and ownership for the final content of this submitted manuscript.

\newpage

\bibliographystyle{IEEEtran}
\bibliography{ref}

\newpage

\appendix

\begin{table*}[t]
    \centering
    \caption{Statistics of the speech training and evaluation datasets sorted by their published years.}
    \label{tab:speech_datasets}
    \resizebox{0.85\textwidth}{!}{
        \begin{tabular}{ccccc}
            \toprule
            \textbf{Dataset} & \textbf{Data Source} & \textbf{Dur.~(hour)} & \textbf{Lang.} & \textbf{Samp. Rate (Hz)} \\
            \midrule
            DAPS & Studio Recording & 67.5 & EN & 44.1k \\
            HQ-TTS & Studio Recording & 191.0 & \makecell{Bangla/Javanese/Khmer \\ Nepali/Sinhala/Sundanese} & 44.1k \\
            AIShell 3 & Studio Recording & 85.0 & ZH & 44.1k \\
            HiFi-TTS & Audio Books & 291.6 & EN & 44.1k \\
            HUI-TTS & Studio Recording & 326.0 & DE & 44.1k \\
            VCTK & Studio Recording & 80.0 & EN & 96k \\
            Bible-TTS & Audio Books & 420.3 & African Languages & 48k \\
            EARS & Studio Recording & 100.0 & EN & 48k \\
            Mana-TTS & Studio Recording & 100.0 & Persian & 44.1k \\
            \midrule
            Common Voice & Studio Recording & 5626.0 & \makecell{EN/ZH/JA \\ KO/FR/DE} & 44.1k \\
            Hitsugi & Studio Recording & 0.4 & JA & 48k \\
            ZunzunProject & Studio Recording & 18.9 & JA & 96k \\
            Voice Seven & Studio Recording & 42.6 & JA & 96k \\
            Coeiroink & Studio Recording & 0.6 & JA & 48k \\
            Amitaro & Studio Recording & 5.9 & JA & 44.1k \\
            Narakuyui & Studio Recording & 0.7 & JA & 48k \\
            Matsukane & Studio Recording & 1.5 & JA & 96k \\
            \bottomrule
        \end{tabular}
    }
\end{table*}

\begin{table*}[t]
    \centering
    \caption{Statistics of the music training and evaluation datasets sorted by their published year.}
    \label{tab:music_datasets}
    \resizebox{0.7\textwidth}{!}{
    \begin{tabular}{cccc}
        \toprule
        \textbf{Dataset} & \textbf{Data Source} & \textbf{Dur.~(hour)} & \textbf{Samp. Rate (Hz)} \\
        \midrule
        GoodSounds & Studio Recording & 28.0 & 44.1k \\
        MedleyDB & Multi-Track Projects & 402.9 & 44.1k \\
        MUSDB18 & Multi-Track Projects & 49.1 & 44.1k \\
        Slakh2100 & Kontact Sound Libraries & 1680.1 & 44.1k \\
        Surge Synth & Synthesizers & 3.4 & 44.1k \\
        Arturia Synth & Synthesizers & 0.7 & 44.1k \\
        DX7 Synth & Synthesizers & 22.5 & 44.1k \\
        MoisesDB & Multi-Track Projects & 156.4 & 44.1k \\
        \midrule
        Cambridge Multi-Track & Multi-Track Projects & 783.1 & 44.1k \\
        Cambridge Unmastered & Multi-Track Projects & 14.4 & 44.1k \\
        Internal Dataset & Sample Packs & / & 44.1k \\
        \bottomrule
    \end{tabular}
    }
\end{table*}

\begin{table*}[t]
    \centering
    \caption{Statistics of the audio training and evaluation datasets sorted by their published year.}
    \label{tab:audio_datasets}
    \resizebox{0.6\textwidth}{!}{
    \begin{tabular}{cccc}
        \toprule
        \textbf{Dataset} & \textbf{Data Source} & \textbf{Dur.~(hour)} & \textbf{Samp. Rate (Hz)} \\
        \midrule
        AudioSet-Strong & In-The-Wild & 296.0 & 44.1k \\
        BBC Effects & In-The-Wild & 232.0 & 44.1k \\
        FreeSound & In-The-Wild & 1283.0 & 44.1k \\
        \midrule
        UrbanSound8K & In-The-Wild & 9.0 & 44.1k \\
        ESC50 & In-The-Wild & 3.0 & 44.1k \\
        MACS & In-The-Wild & 11.0 & 44.1k \\
        Internal Dataset & Sample Packs & / & 44.1k \\
        \bottomrule
    \end{tabular}
    }
\end{table*}

\begin{table*}[t]
    \centering
    \caption{Statistics of the singing voice training and evaluation datasets sorted by their published years.}
    \label{tab:singing_datasets}
    \resizebox{0.95\textwidth}{!}{
        \begin{tabular}{cccccc}
            \toprule
            \textbf{Dataset} & \textbf{Data Source} & \textbf{Dur.~(hour)} & \textbf{Style} & \textbf{Lang.} & \textbf{Samp. Rate (Hz)} \\
            \midrule
            NUS-48E & Studio Recording & 2.8 & Children/Pop & EN & 44.1k \\
            Opera & Studio Recording & 2.6 & Opera & IT/ZH & 44.1k \\
            VocalSet & Studio Recording & 8.8 & Opera & EN & 44.1k \\
            JSUTSong & Studio Recording & 0.4 & Children & JA & 48k\\
            JaCRC
             & Studio Recording & 28.6 & Opera & ZH & 44.1k \\
            PJS & Studio Recording & 0.5 & Pop & JA & 48k \\
            CSD & Studio Recording & 4.6 & Children & EN/KO & 44.1k \\
            JVS-Music & Studio Recording & 30.0 & Children & JA & 48k \\
            KiSing & Studio Recording & 0.9 & Pop & ZH & 44.1k \\
            OpenSinger & Studio Recording & 51.8 & Pop & ZH & 44.1k \\
            NHSS & Studio Recording & 4.1 & Pop & EN & 48k \\
            PopCS & Studio Recording & 5.9 & Pop & ZH & 44.1k \\
            PopBuTFy & Studio Recording & 30.7 & Pop & EN & 44.1k \\
            Opencpop & Studio Recording & 5.2 & Pop & ZH & 44.1k \\
            Internal Dataset & Studio Recording & 5.2 & Pop & ZH & 44.1k \\
            M4Singer & Studio Recording & 29.7 & Pop & ZH & 48k \\
            SingStyle111 & Studio Recording & 12.8 & \makecell{Children/Folk/Jazz \\ Opera/Pop/Rock} & EN/IT/ZH & 44.1k \\
            GOAT & Studio Recording & 4.5 & Opera & ZH & 48k \\
            ACESinger & SVS & 321.8 & Pop & EN/ZH & 48k \\
            SingNet-SP & Sample Pack & 334.3 & \makecell{EDM/Folk/Jazz \\ Opera/Pop/Rap} & \makecell{AR/DE/ES/EN \\ FR/ID/PT/RU \\ ZH/MIS} & 44.1k \\
            \midrule
            GTSinger & Studio Recording & 80.6 & \makecell{Children/Folk/Jazz \\ Opera/Pop/Rock} & \makecell{EN/ZH/JA \\ KO/RU/ES \\ FR/DE/IT} & 44.1k \\
            Kiritan & Studio Recording & 1.2 & Pop & JA & 96k \\
            Namine Ritsu & Studio Recording & 14.4 & Pop & JA & 44.1k \\
            Voice Seven & Studio Recording & 7.2 & Pop & JA & 96k \\
            Oniku Kurumi & Studio Recording & 1.5 & Pop & JA & 96k \\
            Ofutonp & Studio Recording & 1.0 & Children & JA & 96k \\
            Yuuri Natsume & Studio Recording & 1.4 & Pop & JA & 48k \\ 
            Amaboshi Cipher & Studio Recording & 3.1 & Children & JA & 44.1k \\
            \bottomrule
        \end{tabular}
    }
\end{table*}

\end{document}